\documentclass[reprint,
 superscriptaddress,
 nofootinbib,
 amsmath,amssymb,
 aps,
 prd,
 floatfix,
]{revtex4-2}

\usepackage{color}
\usepackage{graphicx}
\usepackage{dcolumn}
\usepackage{bm}
\usepackage[font=normalsize,labelfont=sf,textfont=sf]{subfig}
\usepackage{array}
\usepackage{autobreak}
\usepackage{multirow}
\usepackage{microtype}
\usepackage{balance}
\usepackage[normalem]{ulem}
\usepackage[colorlinks,
            linkcolor=blue,
            urlcolor=blue,
            filecolor=blue,
            anchorcolor=blue,
            citecolor=blue]{hyperref}
\allowdisplaybreaks

\begin{document}

\preprint{APS/123-QED}

\title{Search for $CP$ Violations in the Production and Decay\\ of the Hyperon-Antihyperon Pairs}
\author{Mengjiao Guo}
\affiliation{Institute of Frontier and Interdisciplinary Science, Shandong University, \\Qingdao 266237, China}
\author{Zhe Zhang}
\email{zhangzhe@impcas.ac.cn}
\affiliation{Southern Center for Nuclear-Science Theory (SCNT), Institute of Modern Physics, \\Chinese Academy of Sciences, Huizhou 516000, China}
\author{Ronggang Ping}
\email{pingrg@ihep.ac.cn}
\affiliation{Institute of High Energy Physics, Chinese Academy of Sciences, \\Beijing 100049, China}
\affiliation{University of Chinese Academy of Sciences, Beijing 100049, China}
\author{Jianbin Jiao}
\email{jiaojb@sdu.edu.cn}
\affiliation{Institute of Frontier and Interdisciplinary Science, Shandong University, \\Qingdao 266237, China}

\date{\today}

\begin{abstract}
This study introduces a complete joint angular distribution analysis for the process $e^{+}e^{-}\to J/\psi\to B(\to B_1\pi)\bar{B}(\to \bar{B}_1\pi)$ with polarized beams, where $B$ and $B_{1}$ are hyperons. We consider both transverse and longitudinal beam polarization and analyze $CP$ violation in the production and decay of hyperons. We present the complete Fisher information matrix for the sensitivities of the weak decay parameters $\alpha_{-},~\alpha_{+},~\phi_{\Xi}$ and $\bar{\phi}_{\Xi}$, the $P$-violating term $F_{A}$, and the EDM $d_{B}$ (derived from the $CP$-violating term $H_{T}$), under the BESIII and STCF statistics. We find that polarization enhances the sensitivity of the parameters. Specifically, longitudinal polarization provides a more significant improvement compared to transverse polarization, with the enhancement being more pronounced for single-step decays than for multistep decays. With the expected statistics from the BESIII experiment, the estimated EDM sensitivities for $\Lambda$ and $\Sigma^{+}$ are on the order of $10^{-19}~e~\rm cm$, while those for $\Xi^{-}$ and $\Xi^{0}$ are $10^{-18}~e~\rm cm$. The upcoming STCF experiment is expected to improve these estimates by $1-2$ orders of magnitude. Furthermore, $CP$ violation in hyperon decays originates from weak-phase differences among partial-wave amplitudes. Based on a refined isospin analysis, we emphasize the non-negligible contribution of $\Delta I = 3/2$ amplitudes to $CP$ violation and advocate for increased theoretical attention to this aspect in future studies.

\end{abstract}

\maketitle

\section{Introduction}

If matter and antimatter were created in equal amounts during the big bang~\cite{PhysRev.70.572.2,PhysRev.71.273,Julie2023BigBT}, the dominance of the observed Universe by ordinary matter, with antimatter being extremely rare, remains a profound puzzle. In 1967, A.D. Sakharov~\cite{Sakharov:1967dj} proposed three essential conditions for generating an asymmetry between matter and antimatter, one of which is the charge conjugation ($C$) violation and the combined violation of $C$ and parity ($P$) symmetries, known as $CP$ violation. $CP$ violation was first observed in neutral $K$ mesons decay~\cite{PhysRevLett.13.138}, and later in $B$~\cite{PhysRevLett.87.091801,PhysRevLett.87.091802} and $D$~\cite{PhysRevLett.122.211803} mesons. 

Within the Standard Model, $CP$ violation is governed by two main parameters: the complex phase in the Cabibbo-Kobayashi-Maskawa (CKM) matrix \cite{PhysRevLett.10.531,10.1143/PTP.49.652}, which influences weak interactions, and the quantum chromodynamics (QCD) angle $\bar{\theta}$~\cite{PhysRevLett.37.172}, which relates to $CP$ violation in strong interactions. The experimental upper bound on the neutron electric dipole moment (EDM) constrains the strong $CP$-violating parameter $\bar{\theta}$ to be less than $10^{-10}$~\cite{PhysRevLett.37.172,PhysRevD.19.2227,PhysRevD.95.013002}. This indicates that the strong $CP$ violation is dynamically suppressed within the Standard Model, a phenomenon known as the strong $CP$ problem \cite{PhysRevD.11.3583,Peccei2008,RevModPhys.82.557}. While experimental research on $CP$ violation has aligned with the predictions of the Standard Model, the
observed degree of $CP$ violation is insufficient to fully account for the matter-antimatter asymmetry in the Universe.

\begin{figure}
\includegraphics[width=0.45\textwidth]{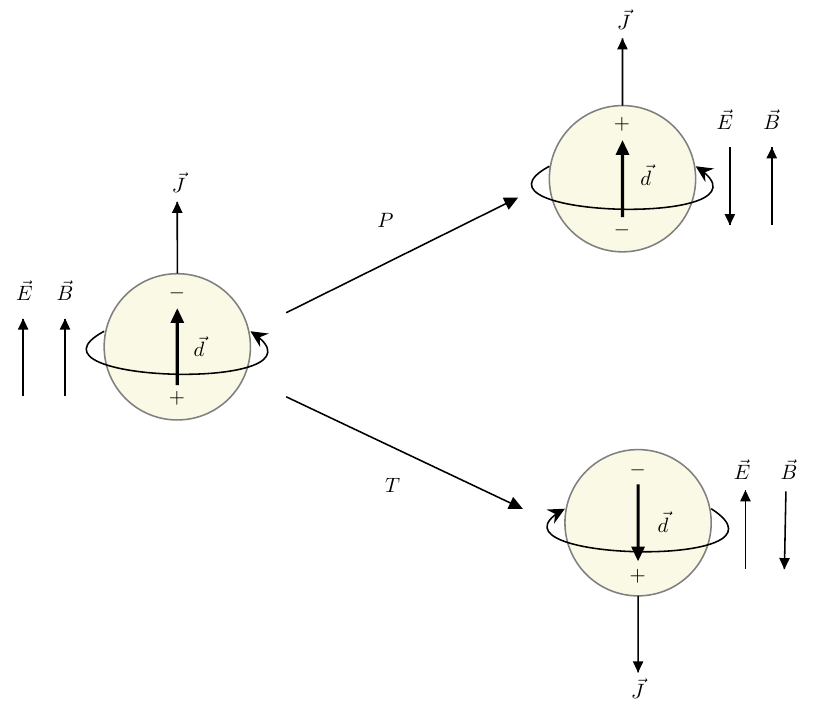}
\caption{\label{fig:EB} Transformations of a spin-$s$ particle under $P$ and $T$ operators, along with the particle's electric dipole moment $(\vec{d})$, and the external magnetic $(\vec{B})$ and electric $(\vec{E})$ fields. A permanent electric dipole moment $\vec{d}$ of a fundamental particle is proportional to its angular momentum $\vec{J}$. It shows that the invariance of the particle dipole moment under $P$ and $T$ transformations is maintained only when $\vec{d}=0$.}
\end{figure}

The search for new physics through $CP$ violation in hyperon production processes has become a promising research direction. In hyperon production, the primary source of $CP$ violation is the EDM. In a magnetic field $\vec{B}$ and electric field $\vec{E}$, a particle’s Hamiltonian can be represented as~\cite{PhysRev.78.807,LANDAU1957127,Bernreuther1991TheED}
\begin{align}
\mathcal{H}=-\vec{\mu}\cdot\vec{B}-\vec{d}\cdot\vec{E}.
\end{align}
Here, $\vec{d}$ and $\vec{\mu}$ denote the EDM and magnetic dipole moment of the particle, respectively. The EDM is a function of charge distribution, $\rho(\vec{r})$,  defined as $\vec{d}=\int \vec{r}\rho(\vec{r})d^{3}r$ and quantifying the electric charge separation. Figure~\ref{fig:EB} illustrates the effect of $P$ and time-reversal ($T$) transformations on a particle with EDM. The EDM of a system $\vec{d}$ must be parallel (or antiparallel) to the average angular momentum of the system $\vec{J}$. The magnetic field $\vec{B}$ and the angular momentum operator $\vec{J}$ are both even under $P$ but odd under $T$, while the electric field $\vec{E}$ is odd under $P$ but even under $T$. Thus, under the constraint of $CPT$ conservation~\cite{PhysRev.82.914,LUDERS19571}, the existence of $\vec{d}\neq0$ implies $C$ and $CP$ violations. 

Since the 1950s, EDM search efforts have yielded constraints on the EDMs of various particles, including leptons \cite{weisskopf1968electric,abdullah1990new,acme2014order,acme2018improved,roussy2023improved,bailey1978new,bennett2009improved,del1990electric,albrecht2000search,inami2003search,ema2022reevaluation}, 
neutrons \cite{smith1957experimental, altarev1979search,pendlebury2015revised,abel2020measurement}, heavy atoms \cite{jacobs1995limit,romalis2001new,griffith2009improved,graner2016reduced}, and protons~\cite{harrison1969experimental,dmitriev2003schiff,PhysRevD.95.013002}. For hyperons, only the EDM of the $\Lambda$, has been measured, with a precision of $|d_{\Lambda}|<1.5\times10^{-16}~e~\rm cm~ (95\% ~C.L.)$ at Fermilab \cite{pondrom1981new}. This upper limit is significantly higher than both the measured limits for nucleons and the theoretical predictions for the $\Lambda$ EDM. Based on neutron EDM constraints, theoretical models predict an upper bound for the $\Lambda$ EDM of approximately $4.4\times10^{-26}~e~\rm{cm}$~\cite{pich1991strong,atwood1992chiral,PhysRevD.61.114017,guo2012baryon}.

In the decay processes of hyperons, $CP$ violation can be investigated through the decay parameters. The decay amplitude of a spin-$1/2$ hyperon into a lighter spin-$1/2$ baryon and a pseudoscalar meson consists of both a $P$-violating S-wave component and a $P$-conserving P-wave component. Consequently, this process can be fully described by two independent decay parameters, $\alpha$ and $\phi$ \cite{PhysRev.108.1645,donoghue1985signals,donoghue1986hyperon}. According to $CP$ symmetry, the decay parameters $\alpha$ and $\phi$ should exhibit opposite signs and equal size between particle and antiparticle observables. For hyperons with multiple strange quarks, the CKM mechanism in the Standard Model predicts very low $CP$ asymmetry values, between $10^{-5}- 10^{-4}$ \cite{PhysRevD.67.056001}. This suggests hyperons are particularly sensitive to $CP$-violating effects beyond the Standard Model, highlighting their importance as probes for new physics. 

The process of $e^{+}e^{-}$ annihilation generates spin-entangled hyperon-antihyperon pairs, offering a unique opportunity to study $CP$ violation in the production and decay processes of hyperons~\cite{PhysRev.178.2288,PhysRevD.76.036005,Perotti:2018wxm,PhysRevD.109.036005,A109,Tomasi-Gustafsson:2005svz,PhysRevD.75.074026,Faldt:2017kgy,PhysRevD.105.116022,Zhang:2024hyq,PhysRevD.108.016011,Zhang:2025oks}. BESIII~\cite{ABLIKIM2010345,Ablikim_2020} has collected the world's largest $J/\psi$ dataset~\cite{Ablikim_2022} and provided the most precise measurements to date of $CP$ violation in hyperon decay processes~\cite{ablikim2022precise,ablikim2020sigma+,PhysRevD.106.L091101,BESIII:2021ypr,PhysRevD.108.L031106,PhysRevD.108.L011101,PhysRevLett.126.092002,Xiao2025}. Theoretical discussions on the EDM of this process have also begun gradually~\cite{PhysRevD.47.R1744,HE2023137834,Du2023}, while the impact of beam polarization on experimental measurements has garnered significant interest~\cite{PhysRevD.110.034034,PhysRevD.105.116022,Zeng:2023wqw,Zhang_2023,PhysRevD.110.014035,Fu:2023ose,PhysRevD.109.036005}. The proposed Super Tau-Charm Factory (STCF) is expected to increase $J/\psi$ statistics by $2$ orders of magnitude~\cite{STCF}, offering new opportunities to explore $CP$ violation in hyperon-related processes.

In this paper, we present a systematic analysis of $CP$ violation in the production and decay processes of hyperons under conditions where the beam possesses both transverse and longitudinal polarization. For the $e^{+}e^{-}\to J/\psi$ process, we investigate the influence of beam polarization on the polarization of $J/\psi$ and present its complete spin density matrix (SDM). For the decay process $J/\psi\to B\bar{B}~(B$ denotes a hyperon), we provide the production density matrix of hyperons, including the $CP$-violating term $H_{T}$ and the $P$-violating term $F_A$. We establish the relationship between $H_T$, $F_A$, and the EDM of hyperons as well as weak coupling angles. Regarding the decay processes of hyperons, we focus on the $\Lambda,~\Sigma^{+},~\Xi^{-}$ and $\Xi^{0}$ hyperons, and present their decay density matrices. Using the decay parameters of hyperons, we define two parameters, $A_{CP}$ and $B_{CP}$, to characterize $CP$ violation~\cite{PhysRev.108.1645,donoghue1985signals,donoghue1986hyperon}.  These parameters are related to the strong and weak phase angles in the decay processes. We provide the estimate for the partial-wave contributions for hadronic decays of $\Lambda$ and $\Xi$ hyperons. Finally, we express the joint angular distribution of all products in terms of the production and decay SDMs of hyperons.

We quantify the $CP$ violation sensitivity through a Fisher information matrix (FIM) analysis of the joint angular distribution. This study provides a comprehensive comparison of hyperon EDM and weak decay parameter sensitivities under various beam polarization scenarios at both BESIII and the proposed STCF, including unpolarized beams, $80\%$ transverse polarization of both electron and positron beams, and $80\%$ longitudinal polarization of the electron beam only. Furthermore, we present curves that explicitly show how the sensitivity of hyperon EDMs and weak decay parameters evolves with varying degrees of transverse and longitudinal beam polarization, thus unveiling the impact of beam polarization.

This paper is organized as follows. In Sec.~\ref{sec:PD}, we provide the analysis of $CP$ violations for the production and decay of the hyperons and the joint angular distribution for these processes. In Sec.~\ref{SS}, we derive the statistical sensitivity of $CP$ violations. A summary is given in Sec.~\ref{Sum}.

\section{Production and Decay Chains of Hyperons}\label{sec:PD}

The process $e^{+}e^{-}\to J/\psi\to B\bar{B}$ serves as a crucial tool for studying the properties of hyperons. In this section, we discuss $CP$ violation in these reactions and present the joint angular distributions of all final particles.

\subsection{\texorpdfstring{$J/\psi$ polarization with polarized beams}{}}

\begin{figure}
\includegraphics[width=0.45\textwidth]{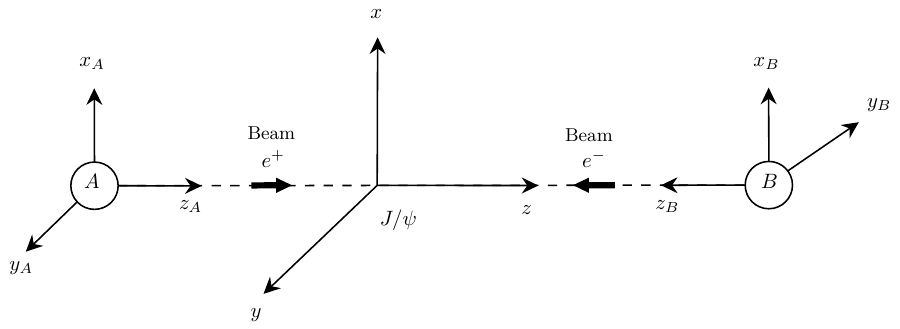}
\caption{\label{beam} Helicity rest frames for electron and positron beams and that for the $J/\psi$ meson in the c.m. frame.}
\end{figure}

A suitable way is to give the density matrices of the electron and positron in their helicity rest frames $S_{A},~S_{B}$, reached from the center-of-mass of the electron-positron pair, as shown in Fig.~\ref{beam}. 

In certain experimental facilities, such as the proposed STCF, longitudinally polarized beams can be prepared, with the longitudinal polarization denoted as $P_L$. Additionally, synchrotron radiation aligns the spins of positrons and electrons either parallel or antiparallel to the magnetic field direction, respectively, leading to transverse polarizations denoted as $P_{T}$. In a electron-positron annihilation experiment with symmetric beam energy, the Sokolov-Ternov effect \cite{Sokolov:1963zn} requires equal degrees of transverse polarization. 

The SDMs of both the leptons in their helicity frame and the $J/\psi$ in the c.m. frame have been extensively discussed in Ref.~\cite{PhysRevD.110.014035},  incorporating both longitudinal and transverse components of the polarization vectors. The SDM of the leptons is represented in their helicity frames as follows\footnote{
Electroweak effects can mimic the angular distributions induced by beam polarization, as noted in Refs.~\cite{Bondar:2019zgm,Zhang:2025oks}. Consequently, the effective beam polarization $P_L^{\prime}$ extracted from experimental fits inherently includes small corrections from electroweak interactions. These corrections, however, are typically very small. More importantly, the angular distributions arising from electroweak effects are distinct from those associated with $CP$ violation. Therefore, our $CP$ violation analysis remains robust and unaffected. A detailed discussion of the electroweak contributions can be found in Refs.~\cite{Bondar:2019zgm,Zhang:2025oks}.}
\begin{align}
&\rho^{-}=\frac{1}{2}\begin{pmatrix}
1+P_{L} & P_{T}\\
P^{*}_{T} & 1-P_{L}\end{pmatrix}~{\rm for}~e^{-},\nonumber\\
&\rho^{+}=\frac{1}{2}\begin{pmatrix}
1+\bar{P}_{L} & P_{T}\\
P^{*}_{T} & 1-\bar{P}_{L}\end{pmatrix}~{\rm for}~e^{+},
\end{align}
taking into account both longitudinal and transverse components of the polarization vectors. In the laboratory system, the SDM element of $J/\psi$ in the $e^{+}e^{-}\to J/\psi$ decay is expressed as
\begin{align}
\rho_{\lambda,\lambda^{\prime}}^{J/\psi}&=2\;\sum_{\lambda_{+},\lambda_{+}^{\prime},\lambda_{-},\lambda_{-}^{\prime}}D^{1*}_{\lambda,\lambda_{+}-\lambda_{-}}(0,0,0)D^{1}_{\lambda^{\prime},\lambda_{+}^{\prime}-\lambda_{-}^{\prime}}(0,0,0)\nonumber\\
&\times \rho^{+}_{\lambda_{+},\lambda_{+}^{\prime}}\rho^{-}_{\lambda_{-},\lambda_{-}^{\prime}}\delta_{\lambda_{+},-\lambda_{-}}\delta_{\lambda_{+}^{\prime},-\lambda_{-}^{\prime}},
\end{align}
where $\lambda_{+}^{(\prime)},\lambda_{-}^{(\prime)}$ represent the helicities of the positron and electron, and $D^{j}_{\lambda^{\prime}\lambda}(0,0,0)$ is the Wigner $D$ matrix. The Dirac $\delta$ function in the above equation ensures the conservation of helicity during the electron-positron annihilation process. For transverse beam polarization, we consider the case where both the electron and positron beams have equal transverse polarization $P_T$. For longitudinal polarization, we focus on the scenario where only the electron beam is longitudinally polarized with polarization $P_L$. 

\subsection{$CP$ violation in hyperon-antihyperon pairs production}

Hyperon-antihyperon pairs produced in $J/\psi$ decays are spin entangled. The production density matrix $R$ can be calculated as
\begin{align}
R_{\lambda_{1},\lambda_{2},\lambda_{1}^{\prime},\lambda_{2}^{\prime}}&\propto \sum_{\lambda,\lambda^{\prime}}\rho^{J/\psi}_{\lambda,\lambda^{\prime}}D^{1*}_{\lambda,\lambda_{1}-\lambda_{2}}(\phi,\theta,0) \nonumber\\
&\times D^{1}_{\lambda^{\prime},\lambda_{1}^{\prime}-\lambda_{2}^{\prime}}(\phi,\theta,0)\mathcal{A}_{\lambda_{1},\lambda_{2}}\mathcal{A}^{*}_{\lambda_{1}^{\prime},\lambda_{2}^{\prime}}.
\end{align}
Here, $\lambda_{1}^{(\prime)}$ and $\lambda_{2}^{(\prime)}$ represent the helicities of the hyperons and their antiparticles. The helicity amplitudes $\mathcal{A}_{\lambda_{1},\lambda_{2}}$ characterize the decay process $J/\psi\to B\bar{B}$, and the angles $\theta$ and $\phi$ denote the hyperon's momentum direction in the $J/\psi$ helicity rest frame. The helicities of the produced hyperons can take the values $(\lambda_{1},\lambda_{2})=(\pm 1/2,\pm 1/2)$, with polarization effects encoded in the SDM. The matrix $R$ thus provides a critical framework for studying polarization-dependent observables in hyperon-antihyperon pairs.

The production density matrix $R$ is physically equivalent to the polarization correlation matrix $S_{\mu\nu}$ of the $B\bar{B}$ system~\cite{PhysRevD.110.014035,Perotti:2018wxm,PhysRevD.105.116022,PhysRevC.31.1857}. The matrix $S_{\mu\nu}$ can be derived from the production density matrix $R$ using the following expression~\cite{PhysRevD.110.034034}:
\begin{align}
S_{\mu\nu}= & \frac{\text{Tr}\left[\left(R_{\lambda_{1},\lambda_{2},\lambda_{1}^{\prime},\lambda_{2}^{\prime}}\right)\left(\Sigma_{\mu}\otimes\Sigma_{\nu}\right)\right]}{\text{Tr}\left[\left(\Sigma_{\mu}\otimes\Sigma_{\nu}\right)\left(\Sigma_{\mu}\otimes\Sigma_{\nu}\right)\right]},
\end{align}
where $\Sigma_{0}=(1/2)\mathbb{I}_2$, $\Sigma_{1}=(1/2)\sigma_x$, $\Sigma_{2}=(1/2)\sigma_y$ and $\Sigma_{3}=(1/2)\sigma_z$ are the polarization projection matrices for the spin-$1/2$ particles.

The general form of the helicity amplitude for the decay process $J/\psi\to B\bar{B}$ is defined as \cite{PhysRevD.47.R1744,HE2023137834}
\begin{align}
\mathcal{A}_{\lambda_{1},\lambda_{2}}&=\epsilon_{\mu}(\lambda_{1}-\lambda_{2})\bar{u}
(\lambda_{1},p_{1})(\gamma^{\mu}F_{V}+\frac{i}{2m}\sigma^{\mu\nu}q_{\nu}H_{\sigma} \nonumber\\
&+\gamma^{\mu}\gamma^{5}F_{A}+\sigma^{\mu\nu}q_{\nu}\gamma^{5}H_{T})\nu(\lambda_{2},p_{2}).
\end{align}
Here, $m$ is the mass of the $B$ hyperon, and $p_{1}$ and $p_{2}$ represent the four momentum of $B$ and $\bar{B}$, respectively, and $\epsilon_{\mu}$ is the polarization vector of $J/\psi$. The form factors $F_{V}$ and $H_{\sigma}$ correspond to the Dirac and Pauli terms, while $F_{A}$ and $H_{T}$ capture $P$-violating and $CP$-violating terms, respectively. All of these terms are complex valued in the timelike region.

The magnetic and electric form factors $G_{1}$ and $G_{2}$ are defined as \cite{Rosenbluth:1950yq, Korner:1976hv}
\begin{align}
G_{1}=F_{V}+H_{\sigma},~G_{2}=G_{1}-\frac{(p_{1}-p_{2})^{2}}{4m^{2}}H_{\sigma}.
\end{align}
The angular asymmetry parameters $\alpha_{J/\psi}$ and $\Delta\Phi$ are related to these form factors as follows: 
\begin{align}
&\alpha_{J/\psi}=\frac{M^{2}_{J/\psi}|G_{1}|^{2}-4m^{2}|G_{2}|^{2}}{M^{2}_{J/\psi}|G_{1}|^{2}+4m^{2}|G_{2}|^{2}},\nonumber\\
&\frac{G_{1}}{G_{2}}=\left|\frac{G_{1}}{G_{2}}\right|e^{-i\Delta\Phi}.
\end{align}

The $P$-violating term $F_{A}$, primarily arising from $Z$-boson exchange, is related to the weak mixing angle $\theta_{W}$ as
\begin{align}
F_{A}\approx -\frac{1}{6}D g_{V}\frac{g^{2}}{4\cos^{2}\theta_{W}}\frac{1-8\sin^{2}\theta_{W}/3}{m_{Z}^{2}},
\end{align}
where $g=e/\sin{\theta_W}$ is the gauge coupling constants with $e=\sqrt{4\pi/137}$, $D=0.8$ is a weighted coupling constant in the light quark flavor SU(3) limit and  $g_V=1.35~\text{GeV}$ represents the coupling strength between charm quark currents and the $J/\psi$. Then, the expected $P$-violating effect is approximately $-1.07 \times 10^{-6}$ \cite{HE2023137834}.

The $CP$-violating term $H_{T}$ is linked to the EDM of the hyperon through~\cite{PhysRevD.47.R1744,PhysRevD.49.4548,HE2023137834}\footnote{
The relation between the effective coupling vertex $H_T$ and the $\Lambda$ EDM is derived under the assumption of a standard electromagnetic $J/\psi$–$\Lambda\bar{\Lambda}$ vertex, following the treatment in Refs.~\cite{PhysRevD.47.R1744,PhysRevD.49.4548,HE2023137834}. While the presence of exotic contributions in the $J/\psi$ vertex could modify the interpretation of the $H_T$ term in terms of the $\Lambda$ EDM, a detailed study of such corrections is beyond the scope of this paper. Importantly, regardless of its microscopic origin, $H_T$ remains a valid and well-defined observable for $CP$ violation.}
\begin{align}
H_{T}=\frac{2e}{3M_{J/\psi}^{2}}g_{V}d_{B}.
\end{align}
Since both  $F_A$  and  $H_T$  are complex in the timelike region, a straightforward parametrization scheme is
\begin{align}
&F_A=|F_A| e^{i \phi_{A}},\\
&H_T=|H_T| e^{i \phi_{T}}.
\end{align}

In subsequent analyses, we treat the magnitude $|F_{A}|$ and phase $\phi_{A}$ as independent fitting parameters, and apply the same approach to $H_T$ by considering its magnitude and phase independently. The magnitude of the EDM in the timelike region,  $|d_B|$, will be used as a reference for comparison with the hadronic EDM in the spacelike region.

Due to the smallness of $P$ and $CP$ violation, contributions from the $F_{A}$ and $H_{T}$ terms to the decay width of $J/\psi$ can be safely neglected such that \cite{Du2023}
\begin{align}
\Gamma_{J/\psi\to B\bar{B}}=
\frac{|G_{1}|^{2}M_{J/\psi}}{12\pi}\sqrt{1-\frac{4m^{2}}{M^{2}_{J/\psi}}}\left(1+\frac{2m^{2}}{M^{2}_{J/\psi}}\left|\frac{G_{2}}{G_{1}}\right|^{2}\right).
\end{align}
When considering $P$ and $CP$ violation effects, along with the longitudinal and transverse polarization of the beams, the production SDM $R$ for the $B\bar{B}$ hyperon pair is described in Appendix~\ref{ProM}.

\subsection{$CP$ violation in hyperon decays}

Due to their short lifetimes, hyperons decay into more stable particles, such as baryons ($B_{1}$) and mesons, through weak interactions. The weak decay severs as the spin self-analyzer. The decay SDM, $T$, for a hyperon decaying into a baryon and a meson is given by
\begin{align}
T_{\lambda_{1},\lambda_{3},\lambda_{1}^{\prime},\lambda_{3}^{\prime}} \propto D_{{\lambda}_{1},\lambda_{3}}^{1/2}(\phi,\theta,0) D^{*1/2}_{\lambda_{1}^{\prime},\lambda^{\prime}_{3}}(\phi,\theta,0)A_{\lambda_{3}} A^{*}_{\lambda_{3}^{\prime}},
\end{align}
where $\lambda_{1}^{(\prime)}$ represents the helicity of the hyperon and $\lambda_{3}^{(\prime)}$ represents the helicity of the baryon in the decay products. The angles $\theta$ and $\phi$ are the polar and azimuthal angles of the produced baryon in the hyperon rest frame. $A_{\lambda}$ denotes the helicity amplitude of this decay process and can be parametrized as~\cite{PhysRev.108.1645,PhysRevD.76.036005,Perotti:2018wxm}
\begin{align}\label{abg}
\alpha_D&=\frac{ |A_{1/2}|^2-|A_{-1/2}|^2}{ |A_{1/2}|^2+|A_{-1/2}|^2}=\frac{2 \text{Re}(A_{S}^{*}A_{P})}{|A_{S}|^{2}+|A_{P}|^2},\nonumber\\
\beta_{D}&=\frac{2 \text{Im}[A_{1/2} A^*_{-1/2}]}{ |A_{1/2}|^2+|A_{-1/2}|^2}=\frac{2 \text{Im}(A_{S}^{*}A_{P})}{|A_{S}|^{2}+|A_{P}|^2},\\
\gamma_{D}&=\frac{2 \text{Re}[A_{1/2} A^*_{-1/2}]}{ |A_{1/2}|^2+|A_{-1/2}|^2}=\frac{|A_{S}|^{2}-|A_{P}|^{2}}{|A_{S}|^{2}+|A_{P}|^2},\nonumber
\end{align}
where $\beta_{D}=\sqrt{1-\alpha_D^2}\sin{\phi_D}$ and $\gamma_{D}=\sqrt{1-\alpha_D^2}\cos{\phi_D}$.  With this parametrization scheme, the specifics of the decay SDM $T$ can be found in Appendix~\ref{deM}. 

The $P$-violating S-wave component and a $P$-conserving P-wave component can be expressed as follows \cite{donoghue1986hyperon}:
\begin{align}
A_{S}&=\sum_{i,j}S_{i,j}e^{i(\delta_{j}^{S}+\phi_{i,j}^{S})} ,\\
A_{P}&=\sum_{i,j}P_{i,j}e^{i(\delta_{j}^{P}+\phi_{i,j}^{P})}.
\end{align}
Here, $ S_{i,j}$  and $P_{i,j}$  are real valued, with  $\{i,j\}=\{2\Delta I,2I\}$  indexing all possible weak isospin transitions and final state isospins. The phase  $\delta_{j}$  represents the phase shift due to strong final-state interactions, while  $\phi_{i,j}$  denotes the $CP$ violation phase arising from weak interactions. For the conjugate process, we have
\begin{align}
\bar{A}_{S}&=-\sum_{i,j}S_{i,j}e^{i(\delta_{j}^{S}-\phi_{i,j}^{S})} ,\\
\bar{A}_{P}&=\sum_{i,j}P_{i,j}e^{i(\delta_{j}^{P}-\phi_{i,j}^{P})}.
\end{align}

In the $CP$ conservation limit, the amplitudes $\bar{A}_S$ and $\bar{A}_P$ for the charge-conjugated decay mode of the antihyperon $\bar{B}$ are $\bar{A}_S=-A_S$ and $\bar{A}_P=A_P$. Therefore, the decay parameters have the opposite values: $\alpha_{D}=-\bar{\alpha}_D$ and $\beta_{D}=-\bar{\beta}_{D}$. Two independent experimental $CP$ violation tests can be defined using these observables~\cite{PhysRev.108.1645,donoghue1985signals,donoghue1986hyperon,PhysRevD.105.116022},
\begin{align}
A_{CP}&:=\frac{\alpha_{D}+\bar{\alpha}_{D}}{\alpha_{D}-\bar{\alpha}_{D}},\label{cp1}\\
B_{CP}&:=\frac{\beta_{D}+\bar{\beta}_{D}}{\alpha_{D}-\bar{\alpha}_{D}}.\label{cp2}
\end{align}
In addition, the following $CP$-violating observable is commonly defined as~\cite{donoghue1985signals,donoghue1986hyperon,PhysRevD.105.116022}
\begin{align}
\Delta_{CP} = \frac{\Gamma - \bar{\Gamma}}{\Gamma + \bar{\Gamma}},
\end{align}
where $\Gamma$ denotes the decay width of the baryon, as detailed in Appendix~\ref{amplitude_anal}.

For convenience, we construct a strong-phase related observable,
\begin{align}
\Delta C=\frac{\beta-\bar{\beta}}{\alpha-\bar{\alpha}}.
\end{align}
The correspondence between these observables and the strong and weak phase angles is listed in Appendix~\ref{de}.

Generally, one believes that the $\Delta I=3/2$ amplitudes are much smaller than the $\Delta I=1/2$ amplitudes, the so-called $\Delta I=1/2$ rule~\cite{Phys.Rev.179.1499,Phys.Rev.Lett.33,Phys.Lett.B.52}. However, recent measurements of $\Lambda$ decay \cite{PhysRevLett.132.101801} have revealed a notable deviation: the ratio of decay asymmetry parameters $\langle\alpha_{\Lambda 0}\rangle/\langle\alpha_{\Lambda -}\rangle$ is  smaller than unity by more than $5\sigma$, providing strong evidence for the presence of a $\Delta I = 3/2$ transition.

\begin{table*}[ht]
\centering
\caption{\label{SandP}Amplitude for the $\Delta I=1/2$ and $\Delta I=3/2$ transitions, and the corresponding $(\Delta I=3/2)/(\Delta I=1/2)$ amplitude ratios, in the $\Lambda$ and $\Xi$ hyperon decays. For each decay mode, the first row presents results without final-state strong phases, while the second row shows amplitudes obtained through Monte Carlo sampling including strong phase effects.}
\renewcommand{\arraystretch}{1.5}
\begin{ruledtabular}
\begin{tabular}{lccccccc}
\multicolumn{2}{c}{}&\multicolumn{2}{c}{$\Delta I=1/2$}&\multicolumn{2}{c}{$\Delta I=3/2$}  &\multicolumn{2}{c}{$(\Delta I=3/2)/(\Delta I=1/2)$}\\\cline{3-4} \cline{5-6} \cline{7-8}
\multicolumn{2}{c}{Decay mode} & $S$ & $P$ & $S$ & $P$ & $S$ ratio & $P$ ratio\\\hline
$\Xi\to \Lambda\pi$&  & $2.0924\pm0.0007$      &  $-0.4011\pm0.0032$    &  $-0.1015\pm0.0008$    & $0.0258\pm0.0040$    &   $-0.0485\pm0.0004$    &  $-0.0644\pm0.0099$   \\\hline
\multirow{3}{*}{$\Lambda\to N\pi$} & Ref.~\cite{PhysRevLett.132.101801} & $\cdot\cdot\cdot$ & $\cdot\cdot\cdot$ & $\cdot\cdot\cdot$ & $\cdot\cdot\cdot$ & $0.0352\pm0.0051$ & $-0.0769\pm0.0095$\\
& $\text{s1}$ & $1.7682\pm0.0072$      & $0.7719\pm0.0085$ & $0.0573\pm0.0075$&$-0.0430\pm0.0078$ &  $0.0324\pm0.0043$ & $-0.0557\pm0.0101$\\   
& $\text{s2}$ & $1.3888\pm0.0069$ & $1.1513\pm0.0087$ & $-0.4792\pm0.0070$ & $0.4936\pm0.0083$ & $-0.3451\pm0.0040$ & $0.4287\pm0.0108$\\\hline
$\Xi\to \Lambda\pi$ &  & $2.0923\pm0.0007$      &  $-0.4013\pm0.0032$    &  $-0.1016\pm0.0008$    & $0.0257\pm0.0040$    &   $-0.0485\pm0.0004$    &  $-0.0641\pm0.0099$   \\\hline
\multirow{2}{*}{$\Lambda\to N\pi$}& $\text{s1}$ & $1.7646\pm0.0072$      & $0.7798\pm0.0086$ & $0.0592\pm0.0077$&$-0.0442\pm0.0079$ &  $0.0336\pm0.0044$ & $-0.0567\pm0.0076$\\   
& $\text{s2}$ & $1.3962\pm0.0070$ & $1.1487\pm0.0088$ & $-0.4729\pm0.0071$ & $0.4840\pm0.0084$ & $-0.3387\pm0.0040$ & $0.4213\pm0.0080$
\end{tabular}
\end{ruledtabular}
\end{table*}

In Appendix~\ref{amplitude_anal}, we present a numerical analysis of the partial-wave amplitudes for $\Lambda$ and $\Xi$ decays. The amplitudes for the $\Delta I=1/2$ and $\Delta I=3/2$ transitions, along with the corresponding $(\Delta I=3/2)/(\Delta I=1/2)$ amplitude ratios, are summarized in Table~\ref{SandP}. For the $\Xi$ hyperon, our results show excellent agreement with those reported in Ref.~\cite{PhysRevD.105.116022}. For the $\Lambda$ hyperon, we identify two distinct solutions that satisfy all imposed constraints. While one solution is consistent with Ref.~\cite{PhysRevD.105.116022}, the other yields a $\Delta I = 3/2$ to $\Delta I = 1/2$ amplitudes ratio exceeding $0.3$. For comparison, the experimentally measured $\Delta I=3/2$ contribution to $\Lambda$ decay~\cite{PhysRevLett.132.101801} is also included in the table. We note that the experimental study presents only the first solution. Although the magnitude of the $\Delta I=3/2$ contribution in that solution is closer to what is observed in other hyperon decay channels, existing measurements do not exclude the possibility of the second solution. Therefore, we emphasize that the exclusive presentation of only the first solution in prior literature may be potentially misleading.

It is important to note that only one of these two solutions is physically admissible. To eliminate the unphysical one, constraints from further experimental observables are needed. Then we examine the weak decay angle $\phi_{\Lambda}$, defined in Eq.~\eqref{abg}

\begin{widetext}
\begin{align}
\sin\phi_{-}=&2\frac{2P_{11}S_{11}\sin(\delta_{11}^{P}-\delta_{11}^{S})-\sqrt{2}P_{11}S_{33}\sin(\delta_{11}^{P}-\delta_{33}^{S})+\sqrt{2}P_{33}S_{11}\sin(\delta_{11}^{S}-\delta_{33}^{P})+P_{33}S_{33}\sin(\delta_{33}^{P}-\delta_{33}^{S})}{\sqrt{1-\alpha_{-}^{2}}\left[2S_{11}^{2}+2P_{11}^{2}+P_{33}^{2}+S_{33}^{2}-2\sqrt{2}P_{11}P_{33}\cos(\delta_{11}^{P}-\delta_{33}^{P})-2\sqrt{2}S_{11}S_{33}\cos(\delta_{11}^{S}-\delta_{33}^{S})\right]},\\
\cos\phi_{-}=&\frac{2S_{11}^{2}-2P_{11}^{2}+S_{33}^{2}-P_{33}^{2}+2\sqrt{2}P_{11}P_{33}\cos(\delta_{1}^{P}-\delta_{3}^{P})-2\sqrt{2}S_{11}S_{33}\cos(\delta_{1}^{S}-\delta_{3}^{S})}{\sqrt{1-\alpha_{-}^{2}}\left[2S_{11}^{2}+2P_{11}^{2}+S_{33}^{2}+P_{33}^{2}-2\sqrt{2}P_{11}P_{33}\cos(\delta_{1}^{P}-\delta_{3}^{P})-2\sqrt{2}S_{11}S_{33}\cos(\delta_{1}^{S}-\delta_{3}^{S})\right]},\\
\sin\phi_{0}=&2\frac{P_{11}S_{11}\sin(\delta_{1}^{P}-\delta_{1}^{S})+\sqrt{2}P_{11}S_{33}\sin(\delta_{1}^{P}-\delta_{3}^{S})-\sqrt{2}P_{33}S_{11}\sin(\delta_{1}^{S}-\delta_{3}^{P})+2P_{33}S_{33}\sin(\delta_{3}^{P}-\delta_{3}^{S})}{\sqrt{1-\alpha_{0}^{2}}\left[S_{11}^{2}+P_{11}^{2}+2S_{33}^{2}+2P_{33}^{2}+2\sqrt{2}P_{11}P_{33}\cos(\delta_{1}^{P}-\delta_{3}^{P})+2\sqrt{2}S_{11}S_{33}\cos(\delta_{1}^{S}-\delta_{3}^{S})\right]},\\
\cos\phi_{0}=&\frac{S_{11}^{2}-P_{11}^{2}+2S_{33}^{2}-2P_{33}^{2}-2\sqrt{2}P_{11}P_{33}\cos(\delta_{1}^{P}-\delta_{3}^{P})+2\sqrt{2}S_{11}S_{33}\cos(\delta_{1}^{S}-\delta_{3}^{S})}{\sqrt{1-\alpha_{0}^{2}}\left[S_{11}^{2}+P_{11}^{2}+2S_{33}^{2}+2P_{33}^{2}+2\sqrt{2}P_{11}P_{33}\cos(\delta_{1}^{P}-\delta_{3}^{P})+2\sqrt{2}S_{11}S_{33}\cos(\delta_{1}^{S}-\delta_{3}^{S})\right]}.
\end{align}
\end{widetext}
Substituting both solutions into this expressions yields clearly separated predictions for $\phi_{\Lambda}$,
\begin{align}
\phi_{-} = &\left\{
\begin{aligned}
&\left(-8.68\pm0.23\right)^{\circ}, \quad \text{solution 1~(s1)}, \\
&\left(-6.00\pm0.20\right)^{\circ}, \quad \text{solution 2~(s2)},
\end{aligned}
\right.\\
\phi_{0} =& \left\{
\begin{aligned}
&\left(-6.19\pm0.19\right)^{\circ}, \quad \text{solution 1~(s1)}, \\
&\left(-196.05\pm0.20\right)^{\circ}, \quad \text{solution 2~(s2)}.
\end{aligned}
\right.
\end{align}
The two solutions yield markedly different predictions for the decay parameter $\phi_\Lambda$ across both decay channels. A particularly striking difference appears in the $\Lambda \to n\pi^0$ decay, where solution 1 predicts $\phi_0 \to 0$, while solution 2 predicts $\phi_0 \to \pi$. To date, no experimental measurement of $\phi_0$ exists for this channel. For $\Lambda \to p\pi^-$, the current world average, $\phi_{-} = -6.5^\circ \pm 3.5^\circ$ \cite{PDG} lacks the precision to discriminate between the solutions. Hence, the primary limitation in determining the magnitude of the $\Delta I=3/2$ contribution in $\Lambda$ decay is the absence of high-precision measurements of $\phi_\Lambda$. Determining $\phi_\Lambda$ requires measuring the polarization of the final-state nucleon, which has been a long-standing experimental challenge. A promising method to achieve this has recently been proposed in Ref.~\cite{liang2025}. Its implementation could resolve the ambiguity in $\phi_\Lambda$ measurements and thereby clarify the size of the $\Delta I=3/2$ amplitude in $\Lambda$ decays.

In previous $CP$ violation studies of hyperon decays\cite{donoghue1985signals,donoghue1986hyperon,He:1991pf,Deshpande:1994vp,Chang:1994wk,He:1995na,He:1999bv,Tandean:2003fr,Tandean:2002vy,He:2025ibc}, the contribution from $\Delta I = 3/2$ partial waves has typically been neglected. This is partly because such amplitudes were generally assumed to be small and difficult to isolate due to the limited experimental precision of earlier measurements. However, more precise experimental determinations of hyperon decay parameters~\cite{PDG} now make the $\Delta I = 3/2$ contribution sufficiently significant to be resolved, as demonstrated in Table~\ref{SandP}.

From a theoretical perspective, calculating the weak phases associated with the $\Delta I = 3/2$ partial waves remains a major challenge in many models. Nonetheless, some frameworks have provided rough estimates~\cite{He:1991pf,Chang:1994wk,He:1995na}. For instance, in left-right symmetric models~\cite{Chang:1994wk}, the weak phase of the $\Delta I = 3/2$ amplitude is predicted to be an order of magnitude larger than that of the $\Delta I = 1/2$ component. If the $\Lambda$ baryon follows the s2 decay scenario, the observed $CP$ violation would primarily stem from the $\Delta I = 3/2$ contribution. Even under the s1 scenario, this component could contribute up to half the effect of the $\Delta I = 1/2$ amplitude. Similar corrections of comparable magnitude may also arise in $\Xi$ decays. It is important to note that theoretical studies of $CP$ violation in hyperon decays still lag significantly behind those in kaon decays (see, e.g., Ref.~\cite{Aebischer:2019mtr}). Our discussion of the $\Delta I = 3/2$ contribution is based on simplified theoretical estimates. We hope that our isospin analysis of hyperon decays will help revitalize interest and progress in this area of research.

\subsection{Joint angular distribution in hyperon production and decay}

\begin{figure}[ht]
\includegraphics[width=0.45\textwidth]{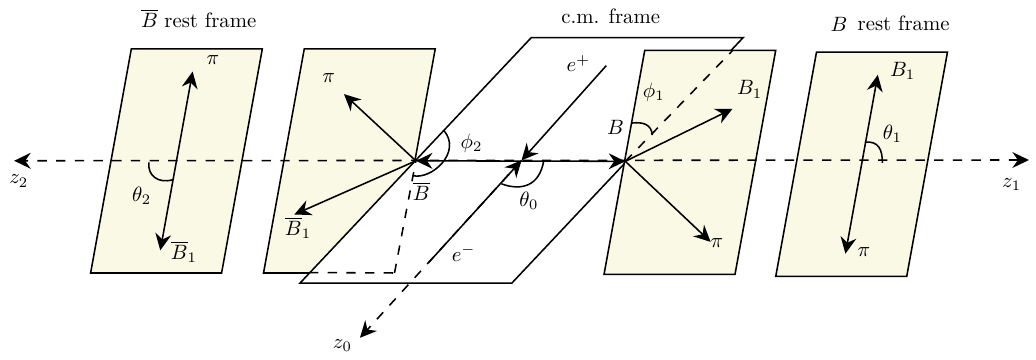}
\caption{\label{helicityL} Helicity angles in single-step decays, such as $J/\psi\to B(\to B_{1}\pi)\bar{B}(\to\bar{B}_{1}\pi)$ decays, where $B$ represents a $\Lambda$ or $\Sigma^{+}$ hyperon.}
\end{figure}

\begin{figure}[ht]
\centering
\includegraphics[width=0.45\textwidth]{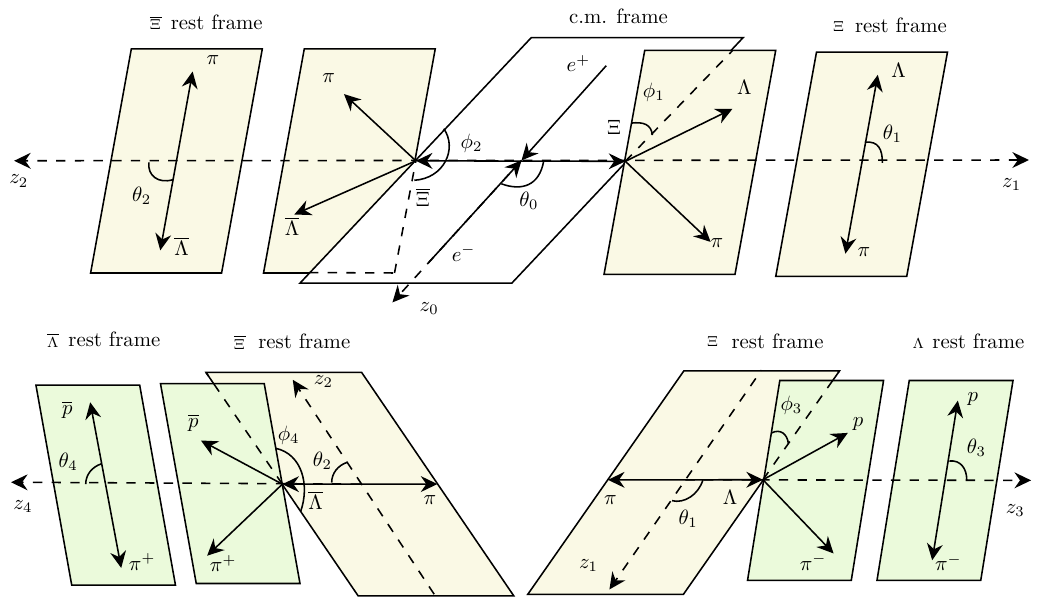}
\caption{\label{helicityXi} Helicity angles in two-step decays, such as $J/\psi\to B(\to B_1\pi)\bar{B}(\to\bar{B}_1\pi)$ decays, where $B$ represents a $\Xi^{0}$ or $\Xi^{-}$ hyperon.}
\end{figure}

\begin{table*}[ht]
\caption{\label{Pub}Measured values of decay parameters for $e^{+}e^{-}\to J/\psi\to B\bar{B}$, with final states listed in the first row and $G_1$ and $G_2$ calculated from $\alpha_{J/\psi}$ and $\Delta\Phi$.}
\begin{ruledtabular}
\renewcommand{\arraystretch}{1.5}
\begin{tabular}{lcccc}
			Decay
channel&\multicolumn{1}{c}{$J/\psi\to\Lambda\bar{\Lambda}$~\cite{ablikim2022precise}}&\multicolumn{1}{c}{$J/\psi\to\Sigma^{+}\bar{\Sigma}^{-}$~\cite{ablikim2020sigma+}}&\multicolumn{1}{c}{$J/\psi\to\Xi^{-}\bar{\Xi}^{+}$~\cite{BESIII:2021ypr}}&\multicolumn{1}{c}{$J/\psi\to\Xi^{0}\bar{\Xi}^{0}$~\cite{PhysRevD.108.L031106}}\\\hline
			 
$\sqrt{s}~(\rm GeV)$           & $M_{J/\psi}$              & $M_{J/\psi}$          & $M_{J/\psi}$          & $M_{J/\psi}$\\
$\alpha^{B}_{J/\psi}$          & $0.4748$         & $-0.508$      & $0.586$       & $0.514$\\
$\alpha_{-}$                   & $0.7519$         & $-0.998$      & $-0.376$      & $-0.375$\\
$\alpha_{+}$             & $-0.7559$        & $0.990$       & $0.371$       & $0.379$\\
$\phi_{\Xi}~(\rm rad)$         & $\cdot\cdot\cdot$& $\cdot\cdot\cdot$& $0.011$       &$0.0051$\\
$\bar{\phi}_{\Xi}~(\rm rad)$   & $\cdot\cdot\cdot$& $\cdot\cdot\cdot$& $-0.021$      &$-0.0053$\\
$\Delta\Phi~(\rm rad)$         & $0.7521$        & $-0.270$      & $1.213$       & $1.168$\\
$G_{1}~(\times 10^{-3})$       & $1.61$                    & $0.86$                & $1.36$                & $1.47$\\
$G_{2}~(\times 10^{-3})$       & $0.97+0.91 i$               & $1.89-0.52 i$           & $0.29+0.76 i$           & $0.38+0.90 i$\\

\end{tabular}
\end{ruledtabular}
\end{table*}

\begin{table*}[ht]
\caption{\label{statistic}Estimated yields of pseudoexperiments based on the statistics from BESIII and STCF
        experiments, where $B_{tag}$ represents the branching ratio of cascade decay,
        $\epsilon_{tag}$ represents the expected detection efficiency, and $N_{tag}^{evt}$
        represents the number of expected events after reconstruction.}
\begin{ruledtabular}
\renewcommand{\arraystretch}{1.5}
\begin{tabular}{lcccc}
			Decay
channel&\multicolumn{1}{c}{$J/\psi\to\Lambda\bar{\Lambda}$}&\multicolumn{1}{c}{$J/\psi\to\Sigma^{+}\bar{\Sigma}^{-}$}&\multicolumn{1}{c}{$J/\psi\to\Xi^{-}\bar{\Xi}^{+}$}&\multicolumn{1}{c}{$J/\psi\to\Xi^{0}\bar{\Xi}^{0}$}\\\hline
			$B_{tag}/(\times10^{-4})$ \cite{PDG}&$7.77$&$2.78$&$3.98$&$4.65$\\		
$\epsilon/\%$ \cite{ablikim2022precise,ablikim2020sigma+,BESIII:2021ypr,PhysRevD.108.L031106}&$40$&$25$&$15$&$7$\\
			$N_{tag}^{evt}/(\times10^{5})$(BESIII) \cite{ablikim2022precise,ablikim2020sigma+,BESIII:2021ypr,PhysRevD.108.L031106}&$31.3$&$7.0$&$6.0$&$3.3$\\
			
$N_{tag}^{evt}/(\times10^{8})$(STCF) \cite{STCF}&$10.6$&$2.4$&$2.0$&$1.1$\\
\end{tabular}
\end{ruledtabular}
\end{table*}

For the productions and decays of $\Lambda$ or $\Sigma$ pairs, they undergo only single-step decay chains, such as $\Lambda \to p\pi^{-}$ or $\Sigma^{+} \to p\pi^{0}$. The joint angular distribution of all final products can be expressed as
\begin{align}
\frac{d\sigma}{d\Omega}\propto&\mathcal{W}(\boldsymbol{\eta};\boldsymbol{\omega})\nonumber\\
=&\sum_{[\lambda]}R_{\lambda_{1},\lambda_{2},\lambda_{1}^{\prime},\lambda_{2}^{\prime}}
T_{\lambda_{1},\lambda_{3},\lambda_{1}^{\prime},\lambda_{3}}
T_{\lambda_{2},\lambda_{4},\lambda_{2}^{\prime},\lambda_{4}}.
\end{align}
The set $[\lambda]$ in the given expression encompasses all helicity symbols participating in the summation, and others.

For the productions and decays of $\Xi^-$ or $\Xi^0$ pairs, they undergo two-step decay chains, such as $\Xi^{-(0)} \to \Lambda \pi^{-(0)}$ followed by $\Lambda \to p \pi^{-}$. The joint angular distribution of all products can be expressed as
\begin{align}
\frac{d\sigma}{d\Omega}\propto&\mathcal{W}(\boldsymbol{\eta};\boldsymbol{\omega})\nonumber\\
=&\sum_{[\lambda]}R_{\lambda_{1},\lambda_{2},\lambda_{1}^{\prime},\lambda_{2}^{\prime}} \nonumber\\
\times& T_{\lambda_{1},\lambda_{3},\lambda_{1}^{\prime},\lambda_{3}^{\prime}}T_{\lambda_{2},\lambda_{4},\lambda_{2}^{\prime},\lambda_{4}^{\prime}}
T_{\lambda_{3},\lambda_{5},\lambda_{3}^{\prime},\lambda_{5}}T_{\lambda_{4},\lambda_{6},\lambda_{4}^{\prime},\lambda_{6}}.
\end{align}
Figure~\ref{helicityL} and Fig.~\ref{helicityXi} illustrate the helicity angles within the corresponding helicity frame for single-step and two-step decays, respectively.

\section{Statistical Significance}\label{SS}

\begin{table*}[ht]
\centering
\caption{\label{Sensitivity}Estimated sensitivities for the weak decay parameters $\alpha_{-}$ (for $\Lambda$, $\Sigma^+$, $\Xi^-$ and $\Xi^0$ decays), $\alpha_{+}$ (for their antiparticle decays), $\phi_{\Xi}$, and $\bar{\phi}_{\Xi}$, the $P$-violating term $|F_{A}|$, and the electric dipole moment $|d_B|$, based on the statistics collected by BESIII and STCF. The statistical quantities corresponding to each decay process used in the calculation under the anticipated conditions of the BESIII and STCF experiments are detailed in Table~\ref{statistic}. The table presents, from top to bottom, the achievable sensitivities in each experiment without polarization, with the addition of $80\%$ longitudinal polarization of the electron beam only, and with the addition of $80\%$ transverse polarization of both electron and positron beams.}
\renewcommand{\arraystretch}{1.5}
\begin{ruledtabular}
\begin{tabular}{lcccccccc}
\multicolumn{1}{c}{}&\multicolumn{2}{c}{$J/\psi\to\Lambda\bar{\Lambda}$}  &\multicolumn{2}{c}{$J/\psi\to\Sigma^{+}\bar{\Sigma}^{-}$}&\multicolumn{2}{c}{$J/\psi\to\Xi^{-}\bar{\Xi}^{+}$}&\multicolumn{2}{c}{$J/\psi\to\Xi^{0}\bar{\Xi}^{0}$}\\\hline
\multicolumn{1}{c}{}& \multicolumn{1}{c}{BESIII} &\multicolumn{1}{c}{STCF} &\multicolumn{1}{c}{BESIII} &\multicolumn{1}{c}{STCF}& \multicolumn{1}{c}{BESIII} &\multicolumn{1}{c}{STCF}& \multicolumn{1}{c}{BESIII} &\multicolumn{1}{c}{STCF}\\\hline

$\alpha_{-}$       & $3.8\times10^{-3}$  & $2.1\times10^{-4}$ & $1.2\times10^{-2}$    & $6.5\times10^{-4}$
                                                              & $2.4\times10^{-3}$    & $1.3\times10^{-4}$
                                                              & $3.2\times10^{-3}$    & $1.8\times10^{-4}$\\
$\alpha_{+}$       & $3.9\times10^{-3}$  & $2.1\times10^{-4}$ & $1.2\times10^{-2}$    & $6.5\times10^{-4}$
                                                              & $2.4\times10^{-3}$& $1.3\times10^{-4}$
                                                              & $3.3\times10^{-3}$    & $1.8\times10^{-4}$\\
$\phi_{\Xi}~(\rm rad)$        & $\cdot\cdot\cdot$& $\cdot\cdot\cdot$& $\cdot\cdot\cdot$& $\cdot\cdot\cdot$& $6.8\times10^{-3}$& $3.7\times10^{-4}$
                                                              & $9.5\times10^{-3}$    & $5.2\times10^{-4}$\\ 
$\bar{\phi}_{\Xi}~(\rm rad)$  & $\cdot\cdot\cdot$& $\cdot\cdot\cdot$& $\cdot\cdot\cdot$& $\cdot\cdot\cdot$& $6.8\times10^{-3}$& $3.7\times10^{-4}$
                                                              & $9.5\times10^{-3}$    & $5.2\times10^{-4}$\\
$|F_{A}|$            & $1.3\times10^{-6}$    & $7.1\times10^{-8}$
                                                              & $2.1\times10^{-6}$    & $1.1\times10^{-7}$
                                                              & $3.1\times10^{-6}$    & $1.7\times10^{-7}$
                                                              & $4.6\times10^{-6}$    & $2.5\times10^{-7}$\\
$|d_B|~(e~\text{cm})$& $8.5\times10^{-19}$   & $4.6\times10^{-20}$
                                                              & $9.7\times10^{-19}$   & $5.3\times10^{-20}$
                                                              & $2.2\times10^{-18}$   & $1.2\times10^{-19}$
                                                              & $3.0\times10^{-18}$   & $1.7\times10^{-19}$  \\\hline
$\alpha_{-}$       & $1.1\times10^{-3}$  & $5.9\times10^{-5}$ & $1.5\times10^{-3}$    & $7.8\times10^{-5}$
                                                              & $2.0\times10^{-3}$    & $1.1\times10^{-4}$
                                                              & $2.8\times10^{-3}$    & $1.5\times10^{-4}$\\
$\alpha_{+}$       & $1.1\times10^{-3}$  & $5.9\times10^{-5}$ & $1.5\times10^{-3}$    & $7.8\times10^{-5}$
                                                              & $2.0\times10^{-3}$    & $1.1\times10^{-4}$
                                                              & $2.8\times10^{-3}$    & $1.5\times10^{-4}$\\
$\phi_{\Xi}~(\rm rad)$        & $\cdot\cdot\cdot$& $\cdot\cdot\cdot$& $\cdot\cdot\cdot$& $\cdot\cdot\cdot$& $4.6\times10^{-3}$    & $2.5\times10^{-4}$
                                                              & $6.3\times10^{-3}$    & $3.4\times10^{-4}$\\ 
$\bar{\phi}_{\Xi}~(\rm rad)$  & $\cdot\cdot\cdot$& $\cdot\cdot\cdot$& $\cdot\cdot\cdot$& $\cdot\cdot\cdot$& $4.6\times10^{-3}$    & $2.5\times10^{-4}$& $6.4\times10^{-3}$    & $3.5\times10^{-4}$\\
$|F_{A}|$          & $9.8\times10^{-7}$      & $5.3\times10^{-8}$
                                                              & $1.8\times10^{-6}$    & $9.7\times10^{-8}$
                                                              & $2.4\times10^{-6}$    & $1.3\times10^{-7}$
                                                              & $3.5\times10^{-6}$    & $1.9\times10^{-7}$\\
$|d_{B}|~(e~\text{cm})$& $4.5\times10^{-19}$& $2.4\times10^{-20}$& $7.9\times10^{-19}$   & $4.3\times10^{-20}$
                                                              & $1.2\times10^{-18}$& $6.8\times10^{-20}$
                                                              & $1.8\times10^{-18}$& $9.8\times10^{-20}$\\\hline
$\alpha_{-}$        & $2.3\times10^{-3}$    & $1.2\times10^{-4}$
                                                              & $6.9\times10^{-3}$    & $3.7\times10^{-4}$
                                                              & $2.2\times10^{-3}$    & $1.2\times10^{-4}$
                                                              & $3.0\times10^{-3}$    & $1.6\times10^{-4}$\\
$\alpha_{+}$        & $2.3\times10^{-3}$    & $1.2\times10^{-4}$
                                                              & $6.9\times10^{-3}$    & $3.7\times10^{-4}$
                                                              & $2.2\times10^{-3}$    & $1.2\times10^{-4}$
                                                              & $3.0\times10^{-3}$    & $1.6\times10^{-4}$\\
$\phi_{\Xi}~(\rm rad)$        & $\cdot\cdot\cdot$& $\cdot\cdot\cdot$& $\cdot\cdot\cdot$& $\cdot\cdot\cdot$& $5.8\times10^{-3}$    & $3.2\times10^{-4}$& $7.5\times10^{-3}$    & $4.1\times10^{-4}$ \\
$\bar{\phi}_{\Xi}~(\rm rad)$  & $\cdot\cdot\cdot$& $\cdot\cdot\cdot$& $\cdot\cdot\cdot$& $\cdot\cdot\cdot$& $5.8\times10^{-3}$    & $3.2\times10^{-4}$& $7.6\times10^{-3}$    & $4.1\times10^{-4}$\\
$|F_{A}|$             & $1.3\times10^{-6}$    & $7.0\times10^{-8}$
                                                              & $1.9\times10^{-6}$    & $1.0\times10^{-7}$
                                                              & $3.1\times10^{-6}$    & $1.7\times10^{-7}$
                                                              & $4.5\times10^{-6}$    & $2.5\times10^{-7}$\\
$|d_{B}|~(e~\text {cm})$& $7.0\times10^{-19}$& $3.8\times10^{-20}$& $8.9\times10^{-19}$   & $4.8\times10^{-20}$
                                                              & $1.8\times10^{-18}$   & $9.9\times10^{-20}$
                                                              & $2.6\times10^{-18}$   & $1.4\times10^{-19}$\\
\end{tabular}
\end{ruledtabular}
\end{table*}

\begin{table*}[ht]
\centering
\caption{\label{CPP}Estimated sensitivities for the $CP$-violating observables $A_{CP}$ and $B_{CP}$ are evaluated using both BESIII and STCF projected statistics, demonstrating significant improvements with the higher luminosity of STCF. The statistical quantities corresponding to each decay process used in the calculation under the anticipated conditions of the BESIII and STCF experiments are detailed in Table~\ref{statistic}. Two distinct computational scenarios are compared: (1) unpolarized beams with full parameter correlations, and (2) $80\%$ longitudinally polarized of the electron beam only with complete parameter correlations.}
\renewcommand{\arraystretch}{1.5}
\begin{ruledtabular}
\begin{tabular}{llcccccc}
\multicolumn{1}{c}{}&\multicolumn{1}{c}{}&\multicolumn{1}{c}{$\delta(A_{CP}^{\Lambda})$}
&\multicolumn{1}{c}{$\delta(A_{CP}^{\Sigma^{+}})$} & \multicolumn{1}{c}{$\delta(A_{CP}^{\Xi^{-}})$}
&\multicolumn{1}{c}{$\delta(B_{CP}^{\Xi^{-}})$}&\multicolumn{1}{c}{$\delta(A_{CP}^{\Xi^0})$}
&\multicolumn{1}{c}{$\delta(B_{CP}^{\Xi^0})$}\\\hline

\multirow{1}{*}{BESIII}& $(1)$& $4.9\times10^{-3}$ & $1.2\times10^{-2}$& $4.5\times10^{-3}$& $1.2\times10^{-2}$& $6.1\times10^{-3}$ & $1.7\times10^{-2}$\\\hline
\multirow{2}{*}{STCF}& $(1)$& $2.7\times10^{-4}$ & $6.5\times10^{-4}$& $2.5\times10^{-4}$& $6.5\times10^{-4}$& $3.4\times10^{-4}$& $9.0\times10^{-4}$\\
                       & $(2)$& $5.9\times10^{-5}$ & $6.4\times10^{-5}$& $2.2\times10^{-4}$& $4.3\times10^{-4}$& $2.9\times10^{-4}$& $6.0\times10^{-4}$\\
\end{tabular}
\end{ruledtabular}
\end{table*}

To assess the sensitivity of the production and decay parameters of hyperons to experimental statistics, we constructed a FIM of these parameters, explicitly accounting for their correlation matrix elements. The joint angular distribution $\mathcal{W}$ is normalized as follows:
\begin{align}
\tilde{\mathcal{W}}=\frac{\mathcal{W}(\bm\eta)} {\int\mathcal{W}(\bm\eta)d\bm\eta},
\end{align}
where ${\bm\eta}=(\theta,\phi,\theta_{1},\phi_{1},\theta_{2},\phi_{2},\theta_{3},\phi_{3},\theta_{4},\phi_{4})$ represents the polar and azimuthal angles of all products. The likelihood function for the observed data is defined as
\begin{align}
L({\bm\eta}|~\bm{\omega})=\prod_{i=1}^{n}\tilde{\mathcal{W}}_{i},
\end{align}
where $n$ is the total number of events, and $\bm{\omega}$ represents the hyperon production and decay parameters. Under the assumptions of the likelihood function satisfies the regularity conditions and the polar angles and azimuthal angles are independent of the parameter $\bm{\omega}$, the FIM element for the maximum likelihood estimate, when the sample size $n$ is sufficiently large, is given by \cite{Fisher1922}
\begin{align}
&I_{ij}({\bm{\omega}}_{i},{\bm{\omega}}_{j})=n\int \frac{1}{\tilde{\mathcal{W}}} \left(\frac{\partial \tilde{\mathcal{W}}}{\partial \bm{\omega}_{i}} \right) \left(\frac{\partial \tilde{\mathcal{W}}}{\partial \bm{\omega}_{j}} \right) d\bm\eta,\nonumber\\
&i,j=1,2,\cdots,k,
\end{align}
where $k$ is the number of parameters. The covariance matrix $V=I^{-1}$ is obtained by inverting the FIM, and the standard error for $\omega_{i}$ is computed as
\begin{align}
\delta(\bm{\omega}_{i}) = \sqrt{V_{ii}}.
\end{align}

\begin{figure*}[ht]
\centering
  \subfloat[$\delta(\alpha_{-})$]{\includegraphics[width=0.5\textwidth]{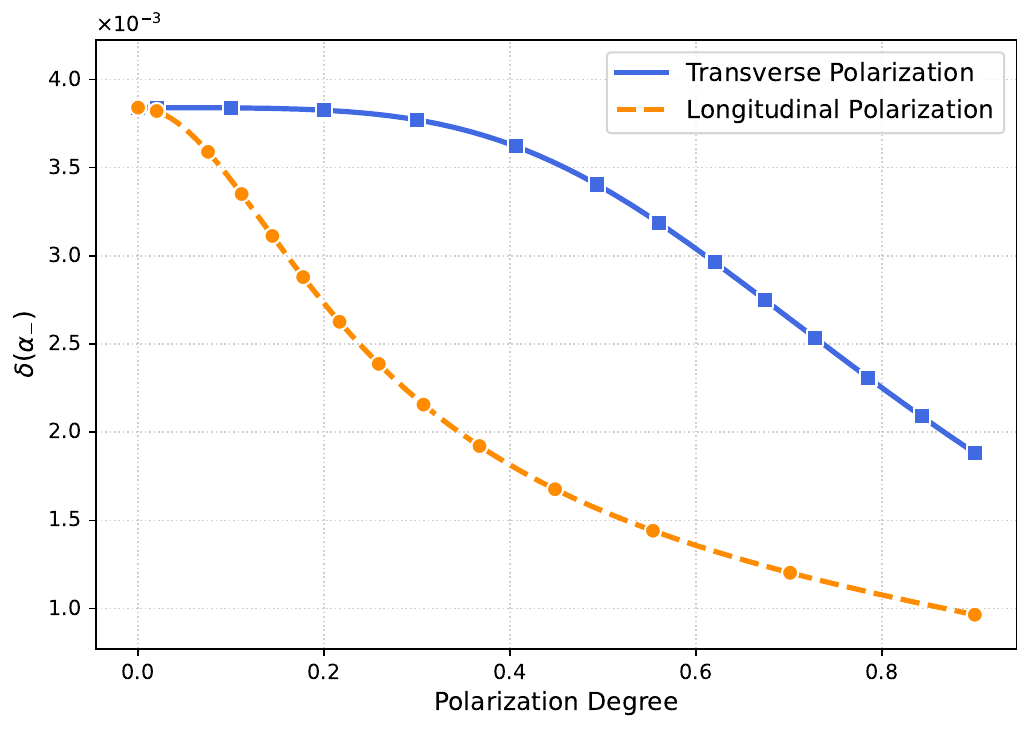}}
 \centering 	
  \subfloat[$\delta(\alpha_{+})$]{\includegraphics[width=0.5\textwidth]{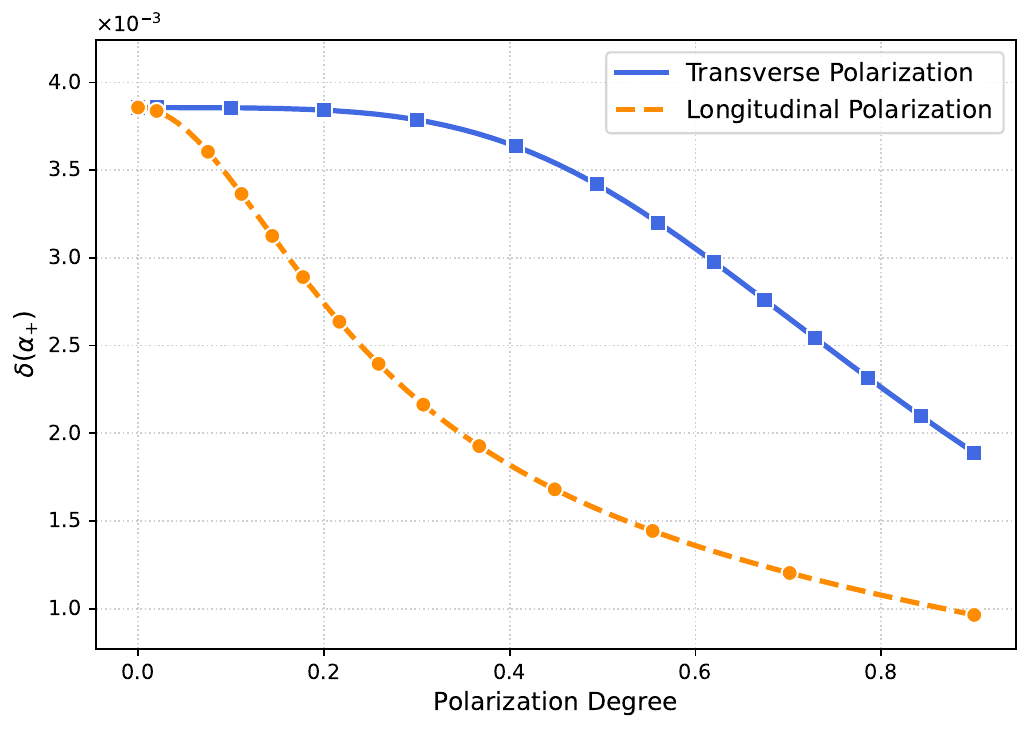}}
 \newline	
  \subfloat[$\delta(|F_{A}|)$]{\includegraphics[width=0.5\textwidth]{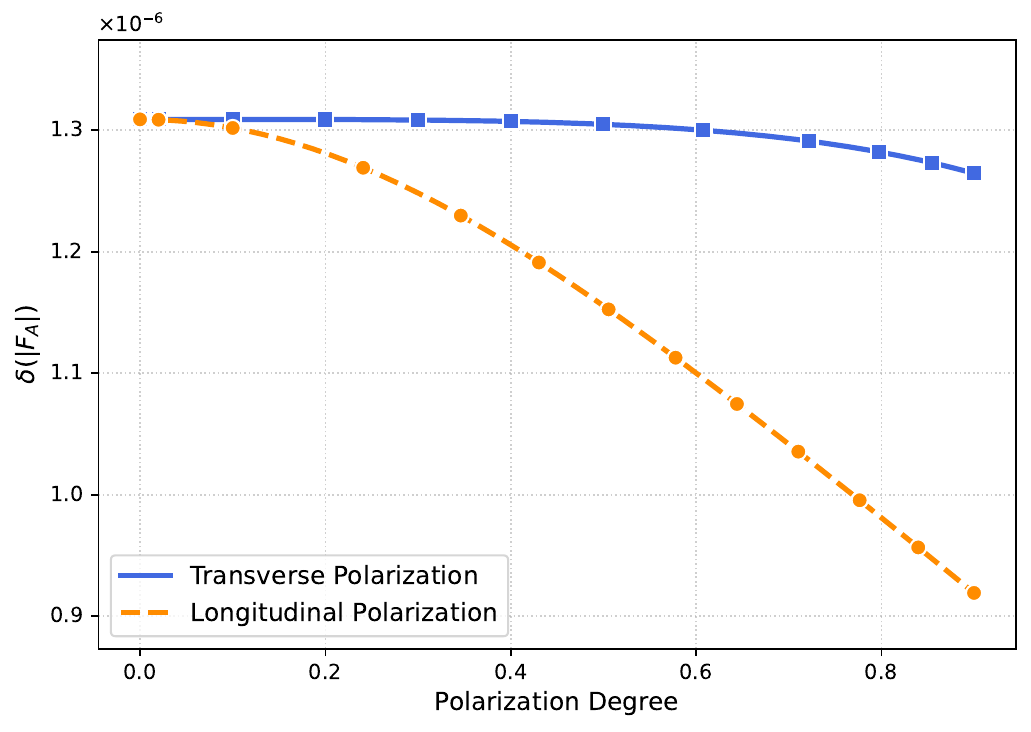}}
  \centering
  \subfloat[$\delta(|d_{\Lambda}|)$]{\includegraphics[width=0.5\textwidth]{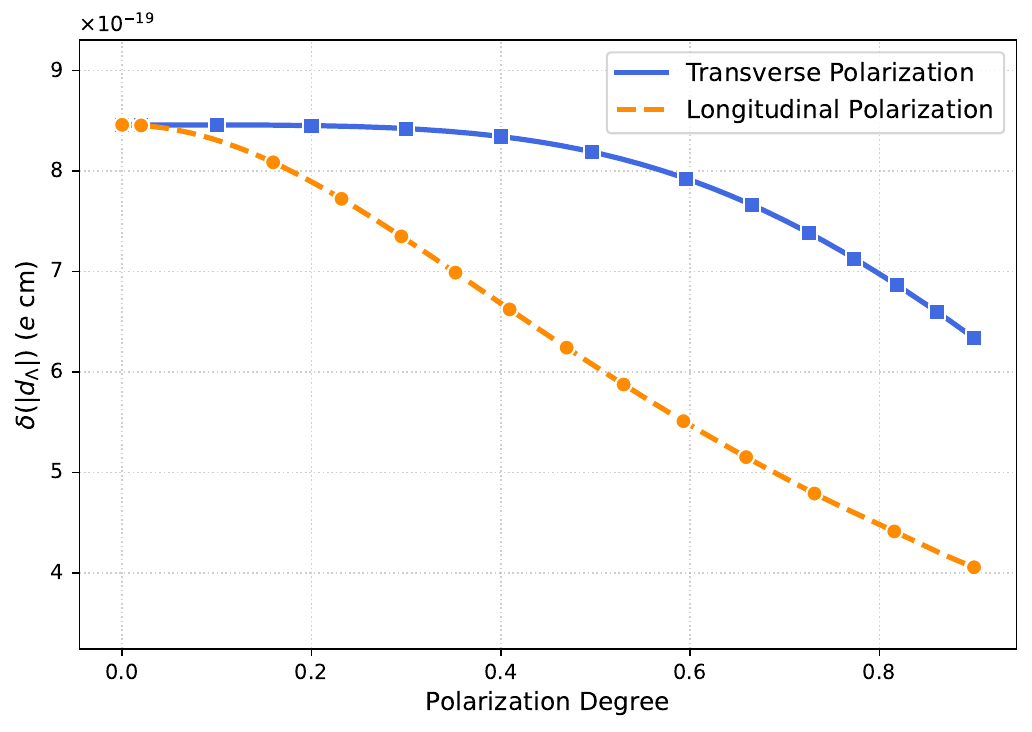}}
\caption{\label{SenLambda}Sensitivities of decay parameters (a) ~$\alpha_{-}$, (b) ~$\alpha_{+}$, (c) the parity-violating term $|F_{A}|$, and (d) the electric dipole moment $|d_{\Lambda}|$ to beam polarization variations in $J/\psi\to\Lambda(\to p\pi^{-})\bar{\Lambda}(\to \bar{p}\pi^{+})$ decays are calculated based on the expected BESIII statistics of $3.13$ million events. The transverse polarization degree refers to the transverse polarization of both the electron and positron beams, while the longitudinal polarization degree corresponds to the longitudinal polarization of the electron beam only.}
\end{figure*}

For $\Lambda$ and $\Sigma^{+}$, only the single-step decay angles $\theta_{1}(\theta_{2})$ and $\phi_{1}(\phi_2)$ are relevant. However, for $\Xi^{-}$ and $\Xi^{0}$, the second-step decay angular variables $\theta_{3}(\theta_4)$ and $\phi_{3}(\phi_4)$ are also incorporated. The hyperon parameter measurements are summarized in Table~\ref{Pub}, and the estimated hyperon statistics collected by BESIII and STCF are shown in  Table~\ref{statistic}.

The sensitivity estimates for the $CP$-violation observables $A_{CP}$ and $B_{CP}$ incorporate correlations among the corresponding decay parameters through rigorous error propagation. The definitions of $A_{CP}$ and $B_{CP}$ are provided in Eqs.~\eqref{cp1} and~\eqref{cp2}. For $A_{CP}$, incorporating correlations between the decay parameters $\alpha_{-}$ and $\alpha_{+}$, the uncertainty is
\begin{align}
\delta(A_{CP})=\sqrt{\sum_{i,j=1}^2 \frac{\partial A_{CP}}{\partial\omega_i}\frac{\partial A_{CP}}{\partial\omega_j}V_{ij}},
\end{align}
where $V_{ij}$ denote the elements of the covariance matrix $V$ for the parameter vector $\bm\omega=(\alpha_-,\alpha_+)$.

\begin{figure*}
\centering
  \subfloat[$\delta(\alpha_{-})$]{\includegraphics[width=0.5\textwidth]{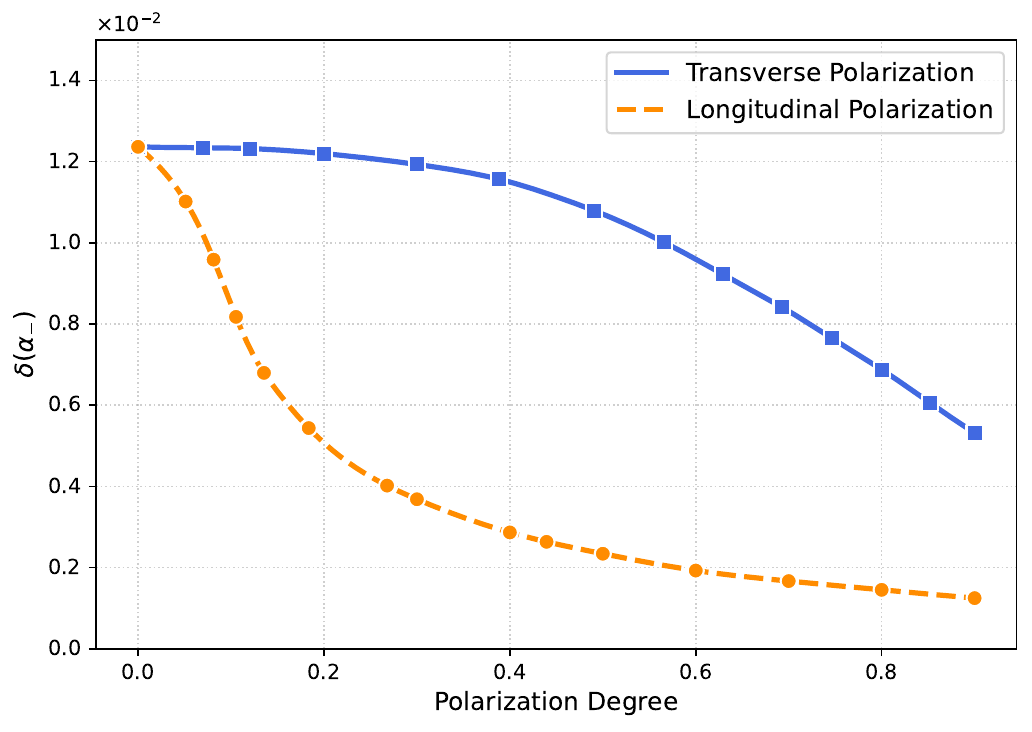}}
 \centering 	
  \subfloat[$\delta(\alpha_{+})$]{\includegraphics[width=0.5\textwidth]{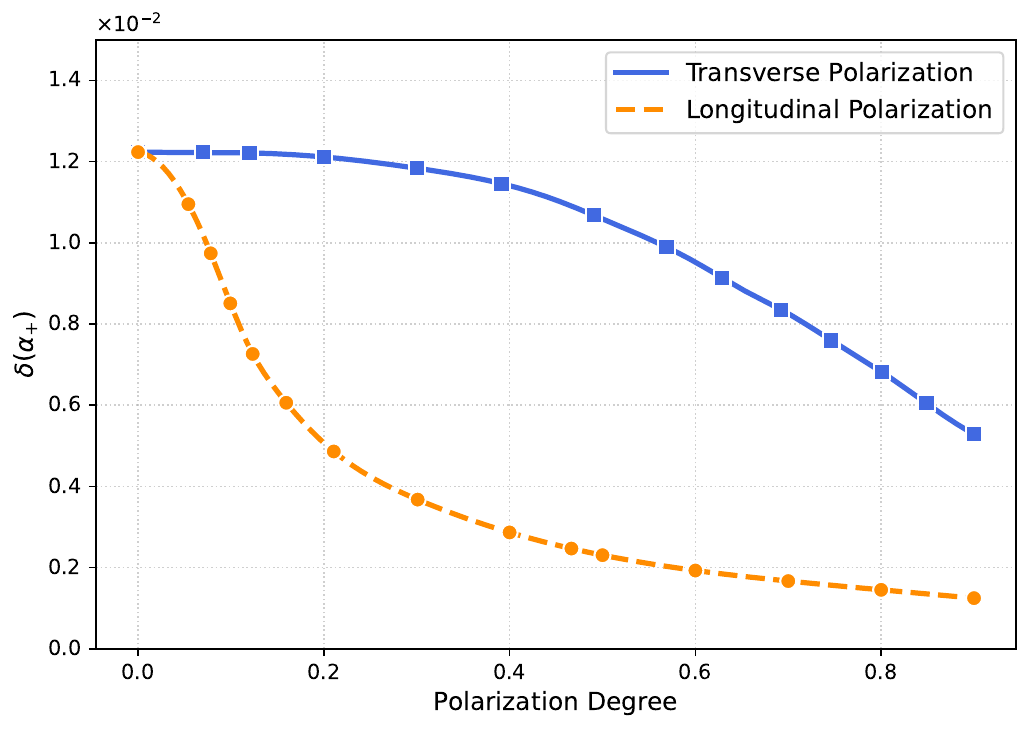}}
 \newline	
  \subfloat[$\delta(|F_{A}|)$]{\includegraphics[width=0.5\textwidth]{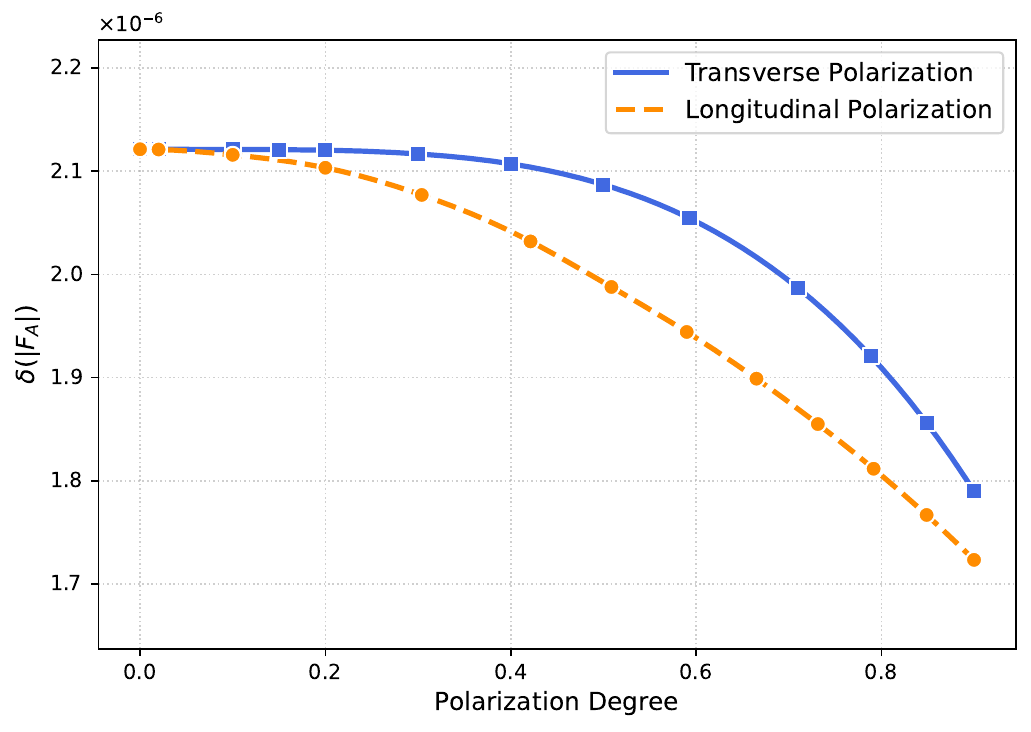}}
  \centering
  \subfloat[$\delta(|d_{\Sigma^{+}}|)$]{\includegraphics[width=0.5\textwidth]{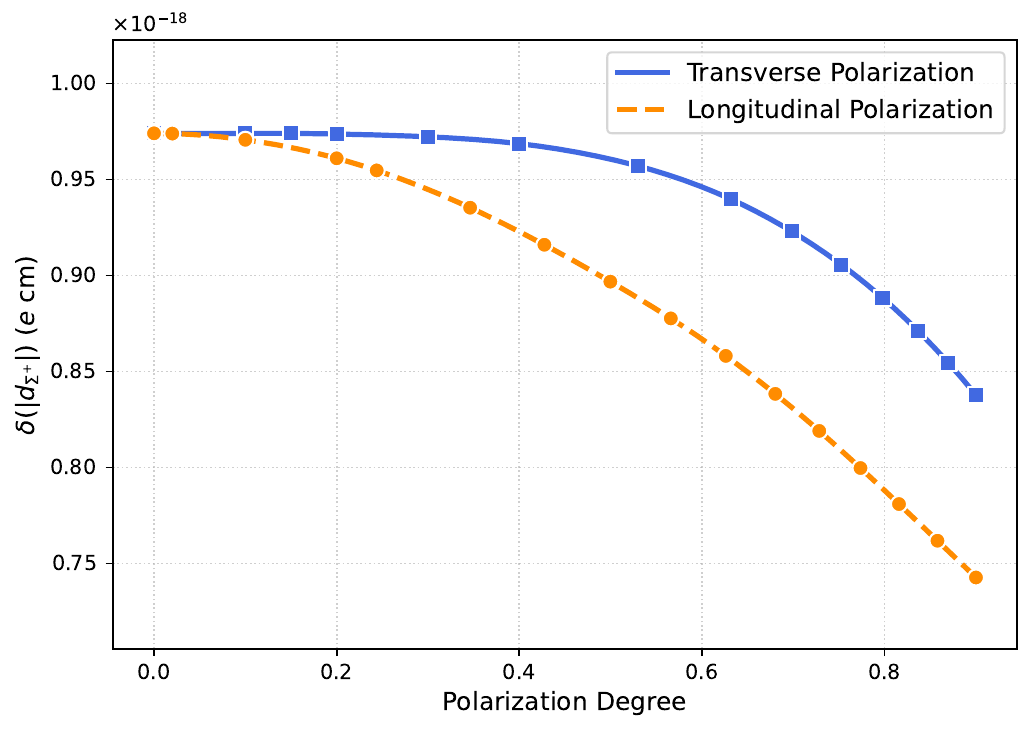}}
\caption{\label{SenSigma}Sensitivities of decay parameters (a)$~\alpha_{-}$ and (b)$~\alpha_{+}$, (c) $P$-violating term $|F_{A}|$, and (d) electric dipole moment $|d_{\Sigma^{+}}|$ to beam polarization variations in $J/\psi\to\Sigma^{+}(\to p\pi^{0})\bar{\Sigma}^{-}(\to\bar{p}\pi^{0})$ decays are calculated based on the expected BESIII statistics of $0.70$ million events. The transverse polarization degree refers to the transverse polarization of both the electron and positron beams, while the longitudinal polarization degree corresponds to the longitudinal polarization of the electron beam only.}
\end{figure*}

\begin{figure*}
\centering
  \subfloat[$\delta(\alpha_{-})$]{\includegraphics[width=0.33\textwidth]{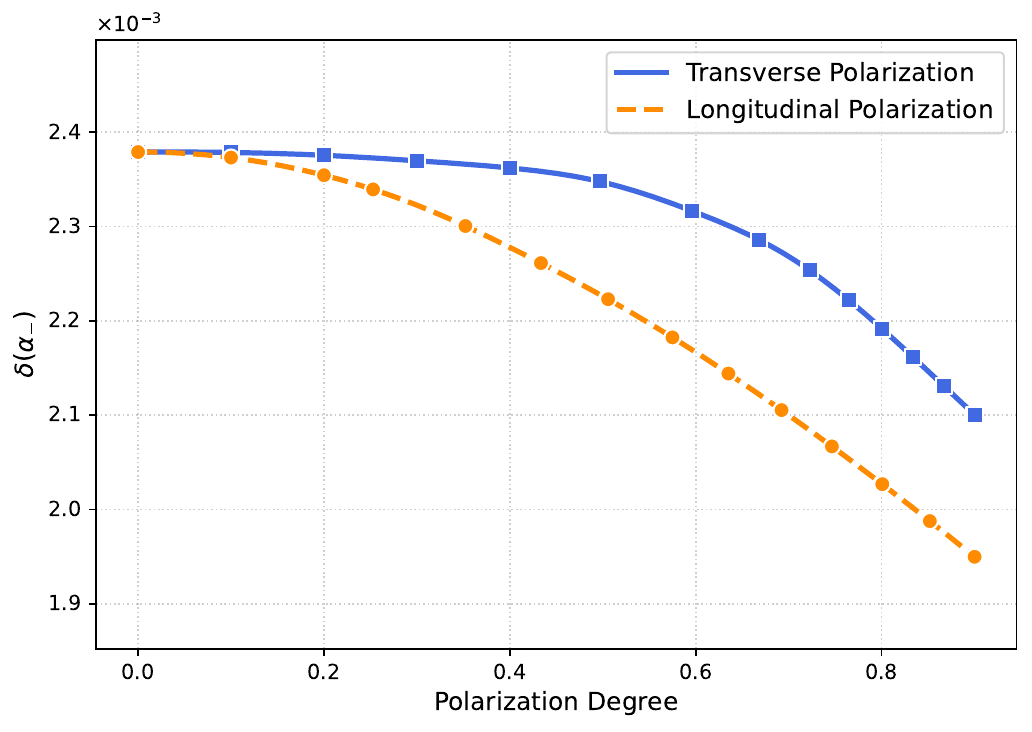}}
 \centering 	
  \subfloat[$\delta(\alpha_{+})$]{\includegraphics[width=0.33\textwidth]{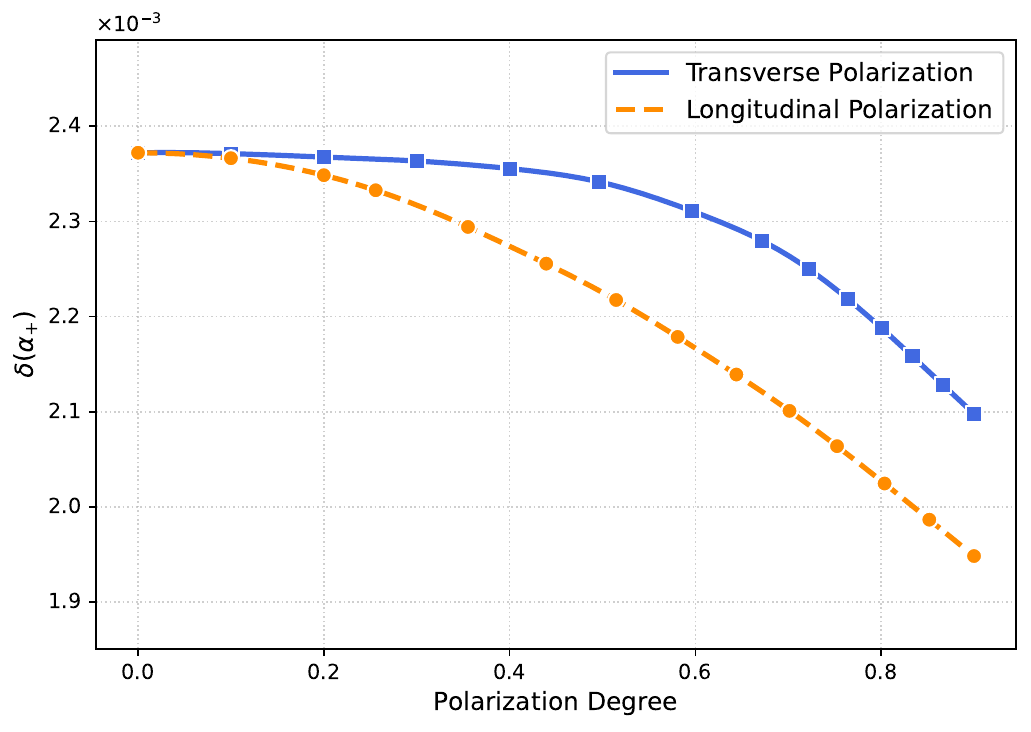}}
 \centering	
  \subfloat[$\delta(\phi_{\Xi})$]{\includegraphics[width=0.33\textwidth]{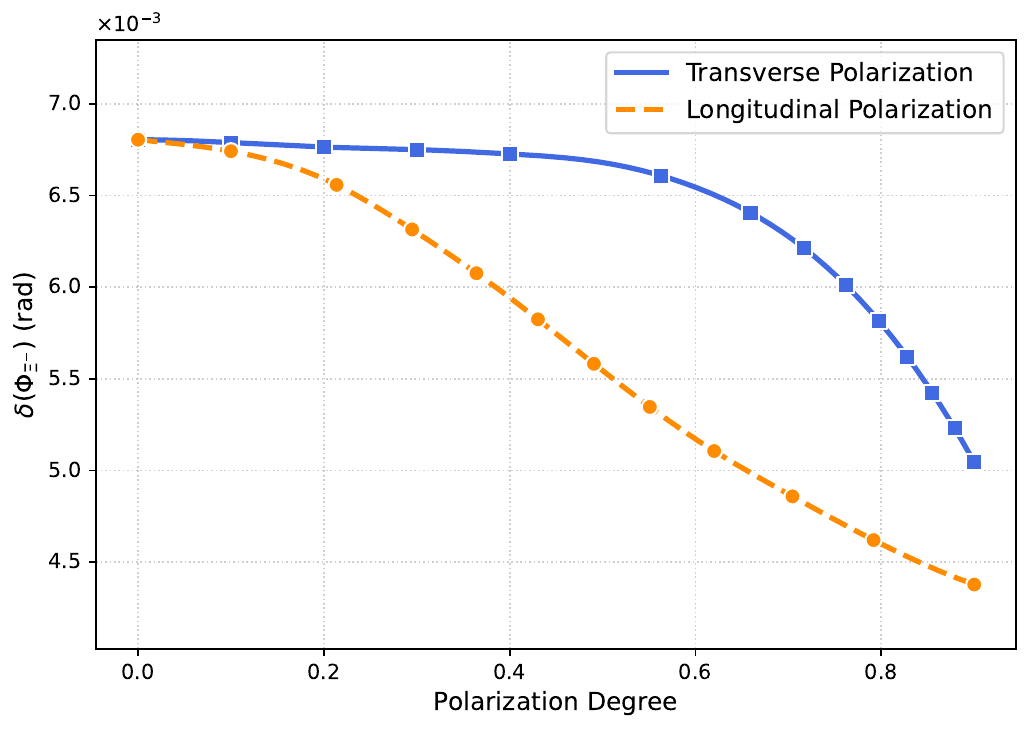}}
 \newline	
  \subfloat[$\delta(\bar{\phi}_{\Xi})$]{\includegraphics[width=0.33\textwidth]{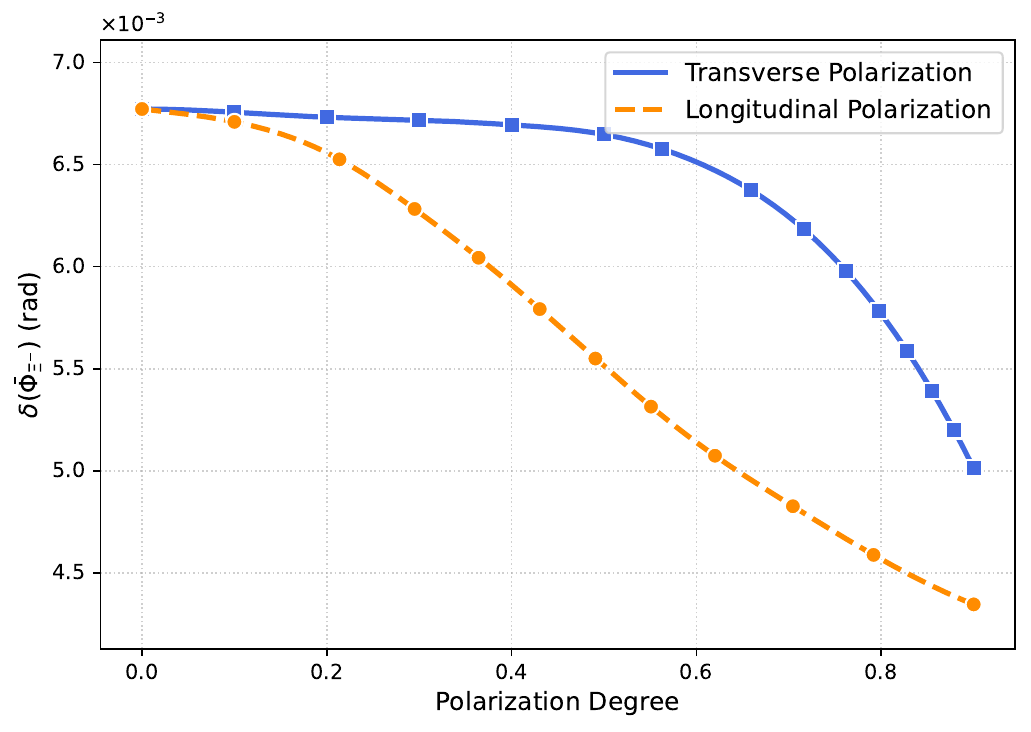}}
  \centering
  \subfloat[$\delta(|F_{A}|)$]{\includegraphics[width=0.33\textwidth]{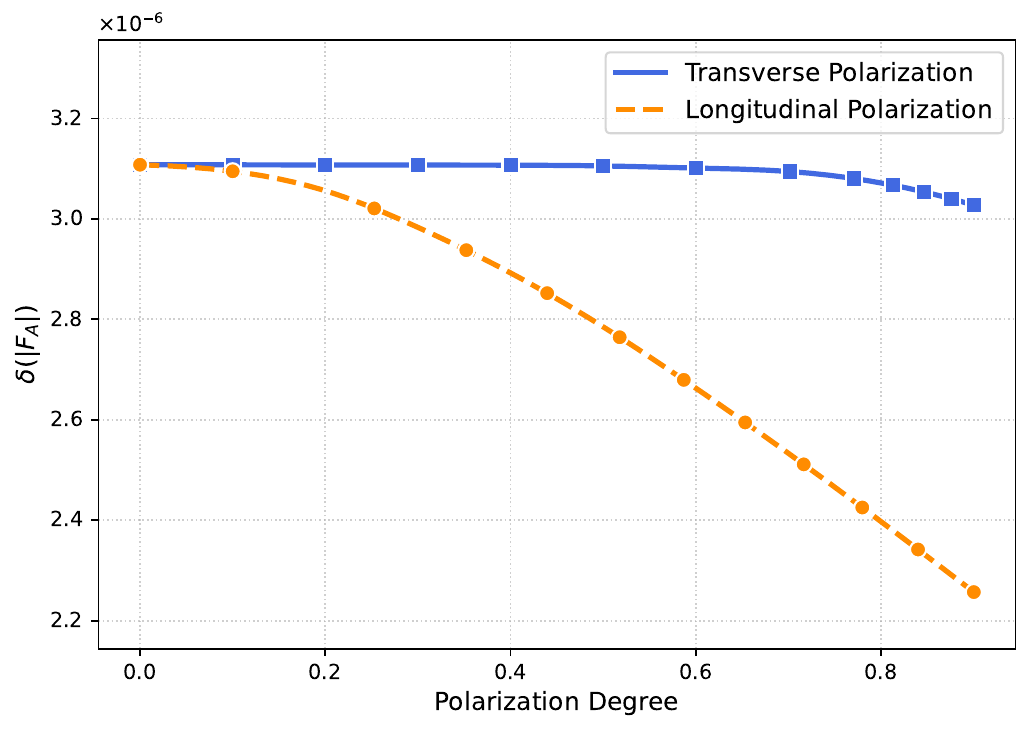}}
  \centering
  \subfloat[$\delta(|d_{\Xi^{-}}|)$]{\includegraphics[width=0.33\textwidth]{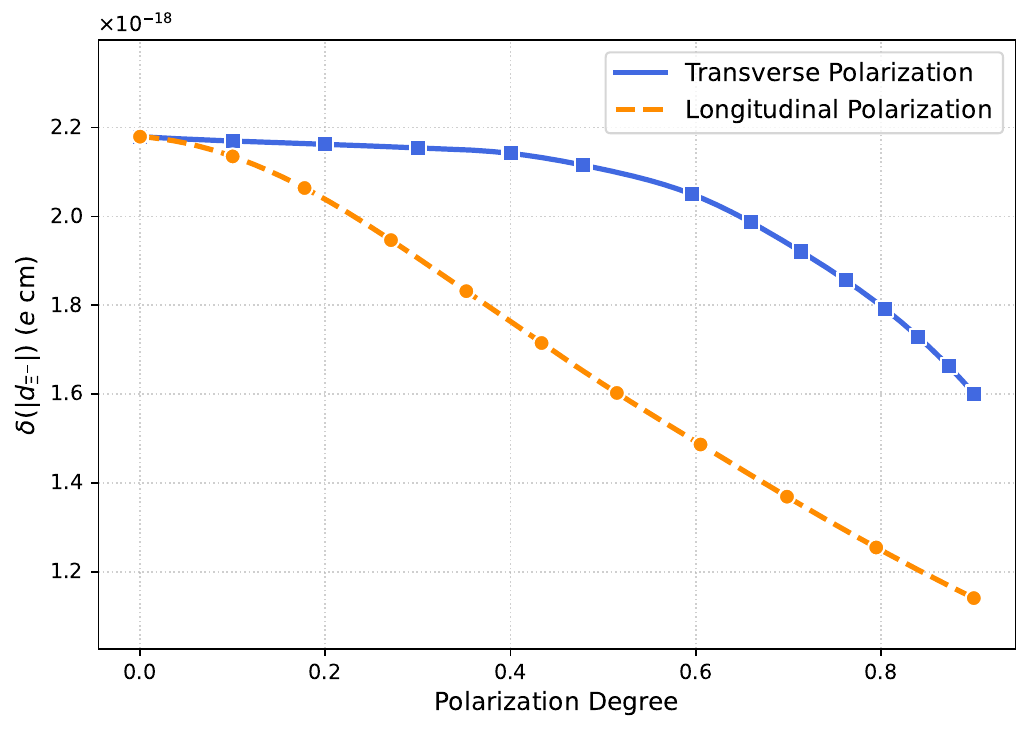}}
\caption{\label{SenXim}Sensitivities for decay parameters (a) $~\alpha_{-}$, (b) $~\alpha_{+}$, (c) $\phi_{\Xi}$, and (d) $\bar{\phi}_{\Xi}$, (e) $P$-violating term $|F_{A}|$, and (d) electric dipole moment $|d_{\Xi^{-}}|$ in $J/\psi\to\Xi^{-}\bar{\Xi}^{+},~\Xi^{-}\to\Lambda(\to p\pi^{-})\pi^{-},~\bar{\Xi}^{+}\to\bar{\Lambda}(\to \bar{p}\pi^{+})\pi^{+}$ decays are calculated based on the expected BESIII statistics of $0.60$ million events. The transverse polarization degree refers to the transverse polarization of both the electron and positron beams, while the longitudinal polarization degree corresponds to the longitudinal polarization of the electron beam only.}
\end{figure*}

\begin{figure*}
\centering
  \subfloat[$\delta(\alpha_{-})$]{\includegraphics[width=0.33\textwidth]{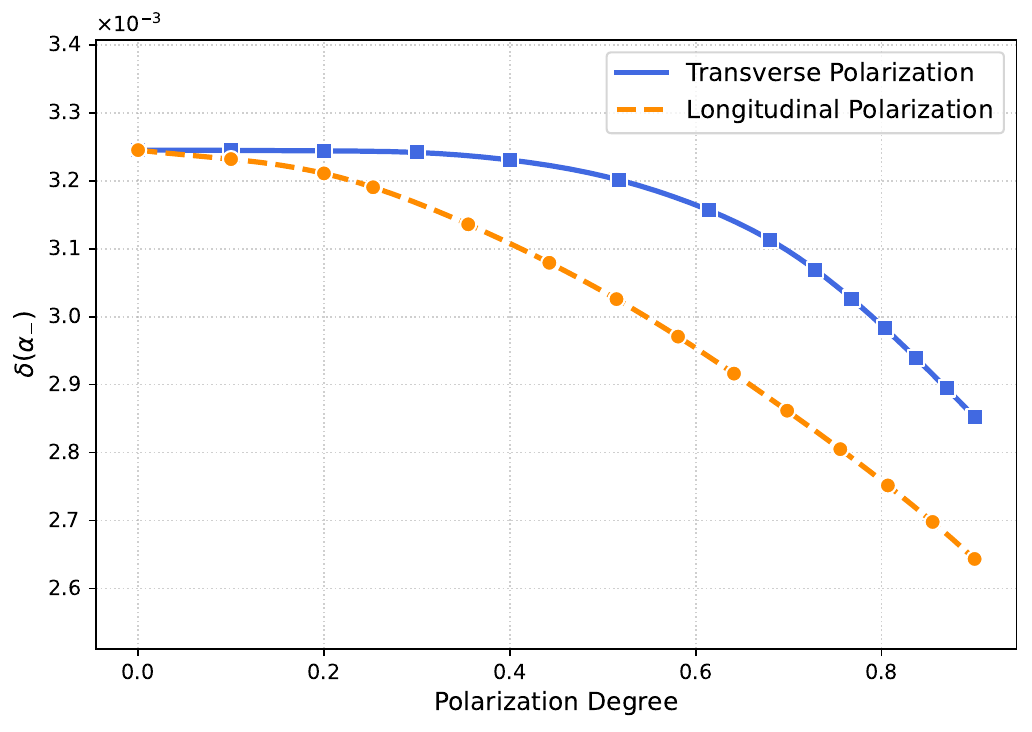}}
 \centering 	 	
  \subfloat[$\delta(\alpha_{+})$]{\includegraphics[width=0.33\textwidth]{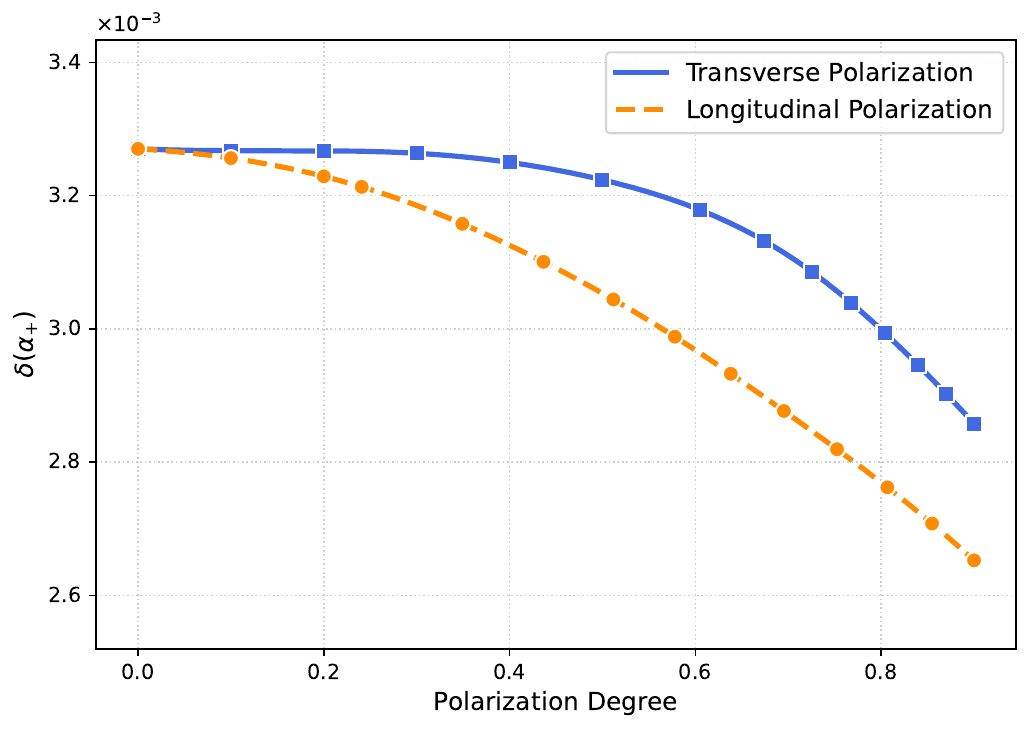}}
 \centering	
  \subfloat[$\delta(\phi_{\Xi})$]{\includegraphics[width=0.33\textwidth]{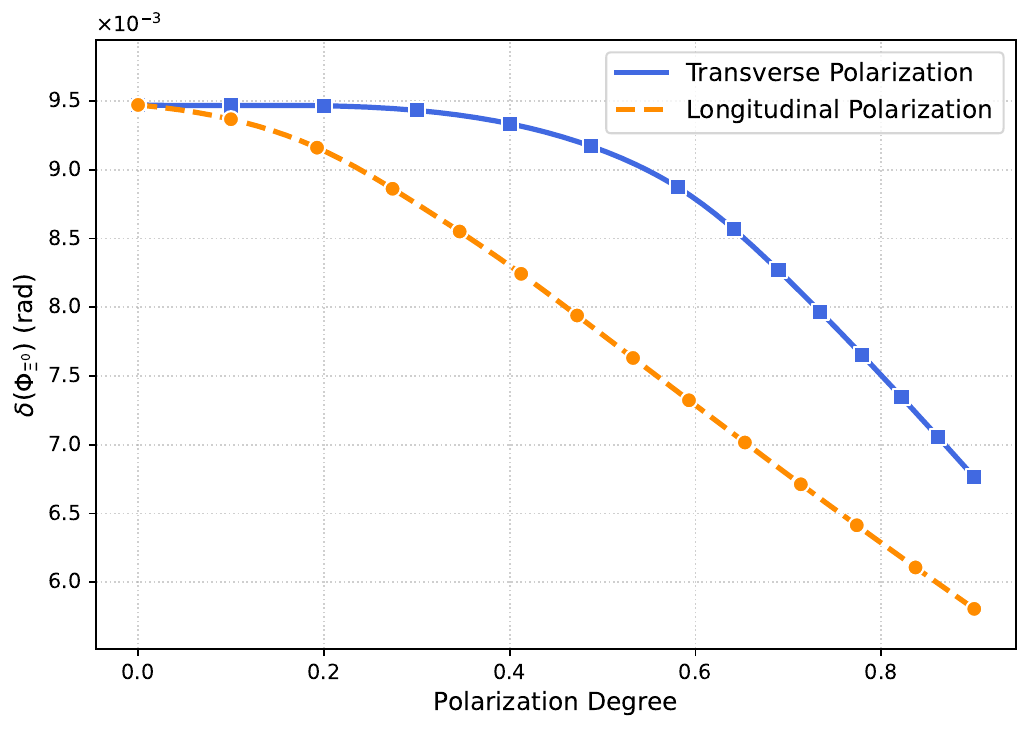}}
 \newline	
  \subfloat[$\delta(\bar{\phi}_{\Xi})$]{\includegraphics[width=0.33\textwidth]{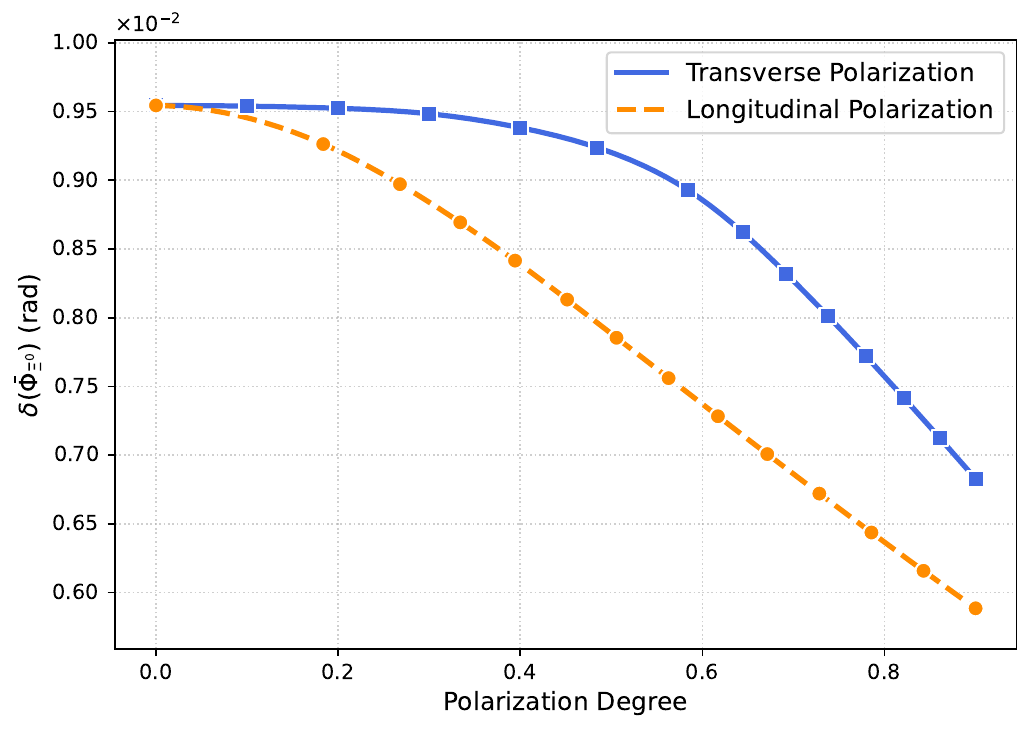}}
  \centering
  \subfloat[$\delta(|F_{A}|)$]{\includegraphics[width=0.33\textwidth]{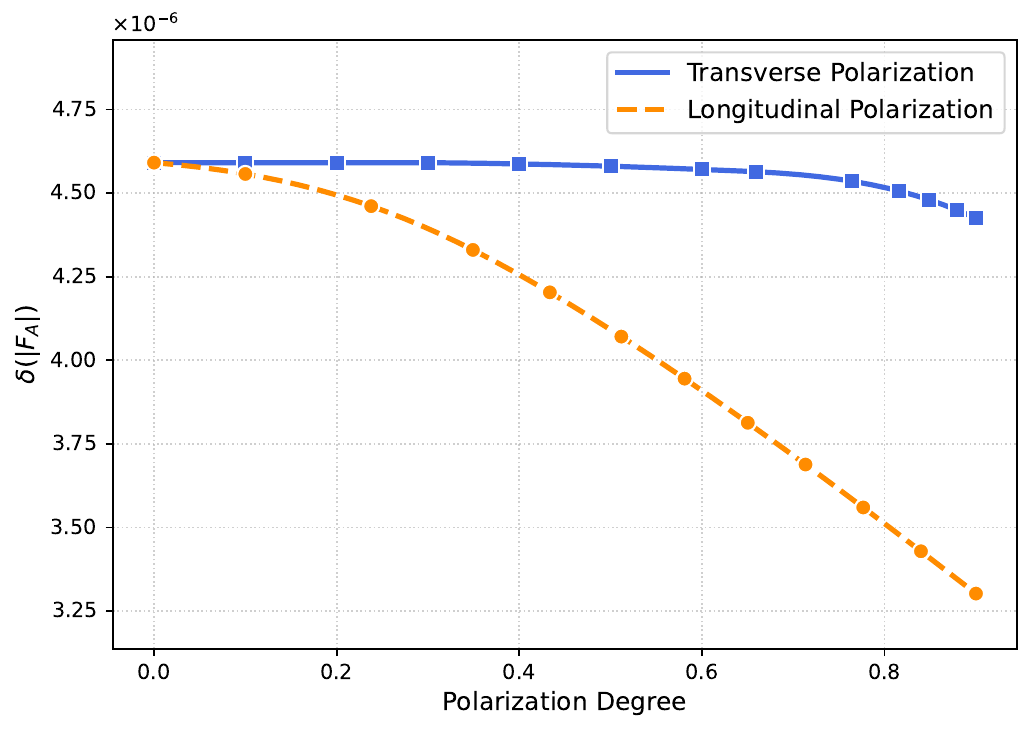}}
  \centering
  \subfloat[$\delta(|d_{\Xi^{0}}|)$]{\includegraphics[width=0.33\textwidth]{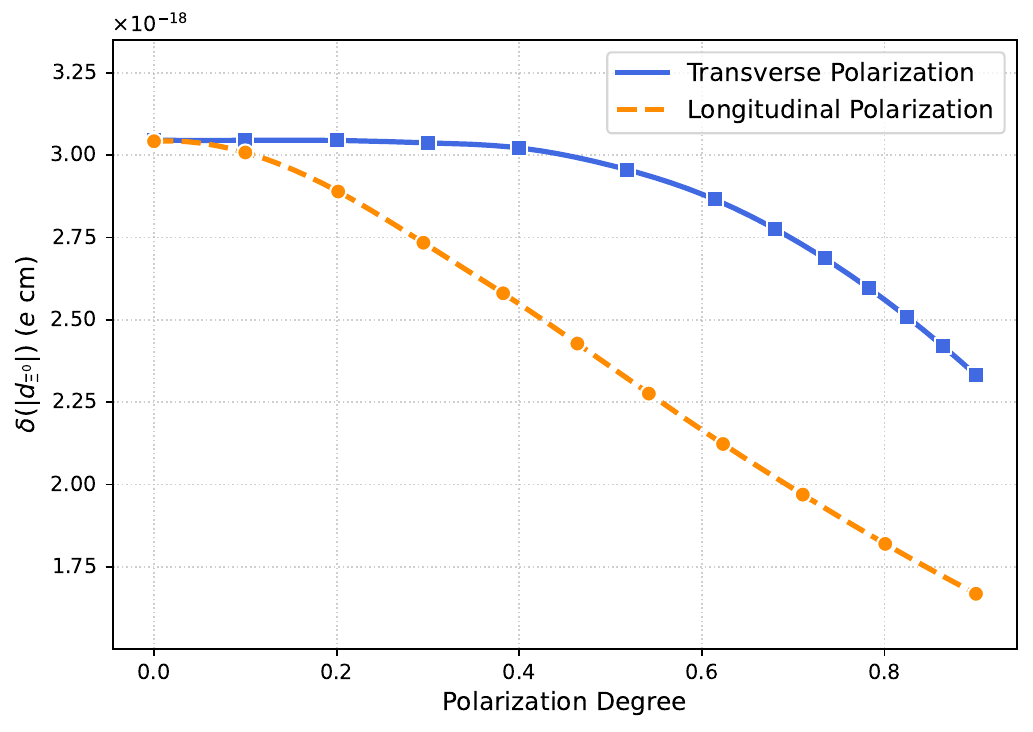}}
\caption{\label{SenXi0}Sensitivities for decay parameters (a) $~\alpha_{-}$, (b) $~\alpha_{+}$, (c) $\phi_{\Xi}$, and (d) $\bar{\phi}_{\Xi}$, (e) $P$-violating term $|F_{A}|$, and (d) electric dipole moment $|d_{\Xi^{0}}|$ in $J/\psi\to\Xi^{0}\bar{\Xi}^{0},~\Xi^{0}\to\Lambda(\to p\pi^{-})\pi^{0},~\bar{\Xi}^{0}\to\bar{\Lambda}(\to\bar{p}\pi^{+})\pi^{0}$ decays are calculated based on the expected BESIII statistics of $0.33$ million events. The transverse polarization degree refers to the transverse polarization of both the electron and positron beams, while the longitudinal polarization degree corresponds to the longitudinal polarization of the electron beam only.}
\end{figure*}

The uncertainty calculation for $B_{CP}$ is more complex due to its dependence on additional parameters $\phi_{\Xi}$ and $\bar{\phi}_{\Xi}$,
\begin{align}
\delta(B_{CP})= \sqrt{\sum_{i,j=1}^4 \frac{\partial B_{CP}}{\partial\omega_i}\frac{\partial B_{CP}}{\partial\omega_j}V_{ij}}.
\end{align}
Here, the covariance matrix $V$ corresponds to the extended parameter vector $\bm\omega=(\alpha_-,\alpha_+,\phi_{\Xi},\bar{\phi}_{\Xi})$.

Previous studies~\cite{PhysRevD.105.116022,PhysRevD.110.014035,Fu:2023ose} have investigated the sensitivity of baryon decay parameters and EDMs under the expected statistics of BESIII or STCF experiments. In this work, we specifically investigate how beam polarization enhances the statistical sensitivity of these parameters. For transverse polarization, both the electron and positron beams must be transversely polarized to induce observable effects in the joint angular distributions of the final-state products. In contrast, for longitudinal polarization, it is sufficient for only one of the beams to be polarized. In this study, we consider the case of a longitudinally polarized electron beam. A detailed discussion can be found in Appendix~\ref{ProM}. 

As shown in Fig.~\ref{SenLambda}, for the $\Lambda$ hyperon, increasing either longitudinal $(P_L)$ and transverse $(P_T)$ polarization enhances the sensitivity of weak decay parameters $\alpha_\pm$ to $\mathcal{O}(10^{-4})$, while improving the $P$-violating term $|F_A|$ to $\mathcal{O}(10^{-7})$ and $|d_\Lambda|$ to $\mathcal{O}(10^{-19})$. Similar enhancement patterns are observed for $\Sigma^{+}$ (Fig.~\ref{SenSigma}), $\Xi^{-}$ (Fig.~\ref{SenXim}), and $\Xi^{0}$ (Fig.~\ref{SenXi0}) hyperons, with $\alpha_\pm$ sensitivities reaching $\mathcal{O}(10^{-3})$, $|F_A|$ achieving $\mathcal{O}(10^{-6})$, and hyperon EDMs $|d_B|$ attaining $\mathcal{O}(10^{-18})$. The $\Xi$ hyperons exhibit additional sensitivity improvements of $\mathcal{O}(10^{-3})$ for both $\phi_\Xi$ and $\bar{\phi}_{\Xi}$.

These results demonstrate that while both longitudinal $(P_L)$ and transverse $(P_T)$ polarization improve the statistical sensitivity. Notably,  polarization effects manifest more strongly in single-step decay baryons $(\Lambda,~\Sigma^+)$ compared to multistep decay baryons $(\Xi)$, and longitudinal polarization provides a much greater improvement compared to transverse polarization. Nevertheless, transverse polarization still yields meaningful sensitivity gains.

Utilizing the projected statistics from both BESIII and STCF experiments, Table~\ref{Sensitivity} presents the estimated sensitivities for the weak decay parameters $\alpha_{-}$, $\alpha_{+}$, $\phi_{\Xi}$, and $\bar{\phi}_{\Xi}$ (only for $\Xi$ hyperons) , the $P$-violating form factor $|F_{A}|$ and the EDM $|d_{B}|$ derived from the $CP$-violating term $|H_{T}|$ under various polarization configurations.

For the BESIII experiment, which currently operates with unpolarized beams, the unpolarized beam scenario provides the most relevant sensitivity estimates. Our sensitivity estimates demonstrate excellent agreement with published BESIII results at comparable statistics~\cite{ablikim2022precise,ablikim2020sigma+,BESIII:2021ypr,PhysRevD.108.L031106}. The current BESIII sensitivities for hyperon decay parameters $\alpha_{\pm}$ reach $\mathcal{O}(10^{-3})$ for $\Lambda,~\Xi^{-}$ and $\Xi^0$, and $\mathcal{O}(10^{-2})$ for $\Sigma^{+}$decays, while the $\Xi$-specific parameters $\phi_{\Xi}$ and $\bar{\phi}_\Xi$ achieve $\mathcal{O}(10^{-3})$ precision. The projected EDM sensitivities are approximately $10^{-19}~e~\rm cm$ for $\Lambda$ and $\Sigma^{+}$ hyperons, representing a $3$-order-of-magnitude improvement for $\Lambda$ EDM compared to the Fermilab measurement~\cite{pondrom1981new}. The $\Xi^{-}$ and $\Xi^{0}$ EDM sensitivities are estimated to be $10^{-18}~e~\rm cm$. As summarized in Table~\ref{CPP}, the achievable precision for $CP$-violating observables $A_{CP}$ is $\mathcal{O}(10^{-3})$ for $\Lambda,~\Xi^{-}$ and $\Xi^0$ decays, and $\mathcal{O}(10^{-2})$ for $\Sigma^{+}$ decays, while $B_{CP}$ for $\Xi^{-}$ and $\Xi^{0}$ can be measurable at $\mathcal{O}(10^{-2})$, showing excellent agreement with previously published BESIII results~\cite{ablikim2022precise,BESIII:2021ypr,PhysRevD.108.L031106}.

The STCF experiment has identified polarized beam operation as a key design requirement, complementing its projected statistics increase of $2-3$ orders of magnitude relative to BESIII. With this higher statistics and an expected $80\%$ longitudinal beam polarization, STCF is anticipated to improve hyperon EDM sensitivities by $1-2$ orders of magnitude beyond BESIII’s unpolarized estimates. Although this improvement remains below the theoretical threshold of $\mathcal{O}(10^{-26})~e~\rm cm$ for hyperon EDMs~\cite{guo2012baryon}, it constitutes a vital advancement in new physics searches. For weak decay parameters, STCF is expected to achieve $\mathcal{O}(10^{-5})$ precision for $\Lambda$ and $\Sigma^+$ decays and $\mathcal{O}(10^{-4})$ for $\Xi^{-}$ and $\Xi^{0}$ decays. The $CP$-violating observables demonstrate particularly promising enhancements: $A_{CP}$ may achieve $\mathcal{O}(10^{-4})$ precision for $\Xi^{-}$ and $\Xi^{0}$ decays while approaching $\mathcal{O}(10^{-5})$ for $\Lambda$ and $\Sigma^+$—approaching the theoretical upper limit of $\mathcal{O}(10^{-5}-10^{-4})$ for hyperon $CP$ violation~\cite{donoghue1985signals,donoghue1986hyperon}. The $B_{CP}$ observable for $\Xi^{-}$ and $\Xi^{0}$ decay can reach $\mathcal{O}(10^{-4})$. These developments, especially the systematic investigation of polarization effects on measurement sensitivity, will offer crucial insights for next-generation collider designs.

\section{Summary}\label{Sum}

This paper systematically analyzes $P$ and $CP$ violation in the production and decay of hyperons in the process $e^{+}e^{-}\to J/\psi\to B(\to B_1\pi)\bar{B}(\to B_1\pi)$. We study the effects of transversely and longitudinally polarized beams on $J/\psi$ polarization. For the decay process $J/\psi\to B\bar{B}$, we derive the production density matrix of hyperon pairs, including the effects of the $P$-violating term $F_A$ and the $CP$-violating term $H_T$. We analyze the correlation between $F_A$ and the weak mixing angle, as well as the connection between $H_T$ and the hyperon EDM. For hyperon decays, we express the decay density matrix in terms of decay parameters and define two $CP$-violating observables, $A_{CP}$ and $B_{CP}$. Finally, we present the joint angular distribution for hyperon production and decay processes.

Using FIM analysis, we evaluate the polarization dependence of statistical sensitivities for weak decay parameters $\alpha$ (including $\phi_{\Xi}$ and $\bar{\phi}_{\Xi}$ for $\Xi$ decays), the $P$-violating term $|F_{A}|$, and hyperon EDMs $|d_B|$, covering $\Lambda$, $\Sigma^{+}$, $\Xi^{-}$, and $\Xi^{0}$ hyperons. The results show that polarization significantly enhances sensitivity for single-step decays compared to multistep decays, with longitudinal polarization providing a much larger improvement than transverse polarization.

We quantitatively estimate the sensitivity of hyperon parameters under BESIII and STCF statistics, and present the parameter sensitivity dependence on beam polarization (including both longitudinal and transverse components) at specified statistics levels. Based on the actual beam polarization capabilities of both BESIII and the proposed STCF experiment, the projected hyperon EDM sensitivities reach $\mathcal{O}(10^{-18})\,e~\text{cm}$ at BESIII, while STCF is projected to reach $\mathcal{O}(10^{-20})\,e~\text{cm}$. For $CP$-violating observables, the estimated precision for $A_{CP}$ reaches $\mathcal{O}(10^{-3})$ at BESIII, while STCF is projected to achieve $\mathcal{O}(10^{-4})$, approaching the theoretical upper limit of $\mathcal{O}(10^{-5}-10^{-4})$ for hyperon $CP$ violation~\cite{donoghue1985signals,donoghue1986hyperon}. The $B_{CP}$ observables in $\Xi^{-}$ and $\Xi^{0}$ decay can reach $\mathcal{O}(10^{-4})$.

The study of $CP$ violation in hyperon decays is tied to the weak phases generated by final-state interactions. These phases originate from the interference among different partial waves (S and P waves) and isospin transitions ($\Delta I=1/2$ and $3/2$). In this work, we perform a comprehensive analysis of hyperon decays by explicitly including the $\Delta I=3/2$ amplitude.  For $\Xi$ decays, our calculations indicate that the $\Delta I=3/2$ contribution is approximately $5\%$ of the dominant $\Delta I=1/2$ amplitude. For the $\Lambda$ hyperon, we identify two viable solutions: a conventional one with a ~$5\%$ contribution and a novel, significantly larger one where the $\Delta I=3/2$ component exceeds $30\%$. Although previous studies have generally overlooked the second solution, we show that it remains consistent with available experimental data. The measurement of the weak decay angle $\phi_\Lambda$ is essential to uniquely determine the physical solution. We demonstrate that the two solutions give distinctly different predictions for $\phi_\Lambda$ in the decay channels $\Lambda \to p\pi^-$ and $\Lambda \to n\pi^0$.  A particularly striking divergence is found in the $\Lambda \to n\pi^0$ channel, where $\phi_\Lambda$ approaches either $0$ or $\pi$. This clear dichotomy offers a well-defined experimental signature. Our results provide new insights into the decay dynamics of the $\Lambda$ hyperon.

We reviewed the theoretical studies of $CP$ violation in hyperon decays and found that, in many models, calculating the weak phases associated with the $\Delta I = 3/2$ amplitudes remains a significant challenge. However, some models predict that the weak phases of the $\Delta I = 3/2$ amplitudes could be more than an order of magnitude larger than those of the $\Delta I = 1/2$ amplitudes~\cite{He:1991pf,Chang:1994wk,He:1995na}. Under such estimates, the contribution of the $\Delta I = 3/2$ partial waves to $CP$ violation could reach half—or even exceed—that of the $\Delta I = 1/2$ components. This highlights the potential importance of higher isospin amplitudes in hyperon decay studies and underscores the need for further theoretical investigation in this direction.

This work provides a theoretical framework for $CP$ violation studies and new physics searches using polarized beams at BESIII and proposed STCF, while offering valuable insights for the design and construction of STCF.

\begin{acknowledgments}
The authors thank Xu Cao, Shengliang Hu, Yutie Liang, Yupeng Pei, Shaojie Wang, Liang Yan, Yong Du, Jianyu Zhang, and Xiaorong Zhou for useful discussions. This work is supported by the National Natural Science Foundation of China (NSFC) under Grants No. 12447132, No. 12375070, No. 12475082, No. 12575112, the National Key R\&D Program of China under Contract No. 2024YFA1611000, No. 2020YFA0406300, the Natural Science Foundation of Shan-dong Province under Contract No. ZR2023MA004, the Joint LargeScale Scientific Facility Funds of the NSFC and CAS under Contract No. U1832207.
\end{acknowledgments}

\section*{DATA AVAILABILITY}
The data that support the findings of this article are openly available \cite{ablikim2022precise,ablikim2020sigma+,BESIII:2021ypr,PhysRevD.108.L031106,PDG,STCF,PhysRevLett.132.101801,HOFERICHTER20161}.

\appendix
\begin{widetext}
\section{PRODUCTION SDM OF HYPERON ANTIHYPERON PAIRS}\label{ProM}

For the production of hyperon–antihyperon pairs, we include both $P$  and $CP$ violating effects in the production mechanism. We also account for the effects of both transverse and longitudinal beam polarizations. For transverse polarization, we consider the case where the electron and positron beams are equally transversely polarized with a polarization degree $P_T$. When only one beam is transversely polarized, no observable effect arises in the production process. For longitudinal polarization, a single polarized beam is sufficient to influence the production. In this work, we consider longitudinal polarization of the electron beam, denoted as $P_L$. If instead the positron beam is longitudinally polarized (denoted $\bar{P}_L$), the relevant expressions can be obtained by the replacement $P_L \rightarrow -\bar{P}_L$. Taking all of the above effects into account, the production SDM $R$ for the hyperon-antihyperon pair can be expressed as

\begin{align}
R_{++++}=&2\sin^{2}\theta(1-P_{T}^{2}\cos 2\phi)[m^{2}|F_{V}|^{2}+\frac{E_{c}^{4}}{{m}^{2}}|H_{\sigma}|^{2}\nonumber\\
-&4m^{2}E_{c}^{2}|H_{T}|^{2}+4 E_{c}^{4}|H_{T}|^{2}+2E_{c}^{2}{\rm Re}[F_{V}H_{\sigma}^{*}]+4 m E_{c}\sqrt{E_{c}^{2}-m^{2}}{\rm Im}[F_{V}H_{T}^{*}]+\frac{4E_{c}^{3}\sqrt{E_{c}^{2}-m^{2}}}{m}{\rm Im}[H_{\sigma}H_{T}^{*}]],\nonumber\\

R_{-+++}=&2\sin\theta[\cos\theta+P_{L}-P_{T}^{2}(\cos\theta\cos2\phi-i\sin2\phi)]\nonumber\\
\times&[mE_{c}|F_{V}|^{2}+\frac{E_{c}^{3}}{m}|H_{\sigma}|^{2}+mE_{c}H_{\sigma}F_{V}^{*}+\frac{E_{c}^{3}}{m}F_{V}H_{\sigma}^{*}-m\sqrt{E_{c}^{2}-m^{2}}F_{A}F_{V}^{*}\nonumber\\
-&2iE_{c}^{2}\sqrt{E_{c}^{2}-m^{2}}F_{V}H_{T}^{*}-2iE_{c}^{2}\sqrt{E_{c}^{2}-m^{2}}H_{\sigma}H_{T}^{*}-\frac{E_{c}^{2} \sqrt{E_{c}^{2}-m^{2}}}{m}F_{A}H_{\sigma}^{*}+2iE_{c}(E_{c}^{2}-m^{2})F_{A}H_{T}^{*}],\nonumber\\

R_{+-++}=&2\sin\theta[\cos\theta-P_{L}-P_{T}^{2}(\cos\theta\cos2\phi+i\sin2\phi)]\nonumber\\
\times&[-mE_{c}|F_{V}|^{2}-\frac{E_{c}^{3}}{m}|H_{\sigma}|^{2}-\frac{E_{c}^{3}}{m}F_{V}H_{\sigma}^{*}-mE_{c}H_{\sigma}F_{V}^{*}-m\sqrt{E_{c}^{2}-m^{2}}F_{A}F_{V}^{*}\nonumber\\
-&\frac{E_{c}^{2}\sqrt{E_{c}^{2}-m^{2}}}{m}F_{A}H_{\sigma}^{*}+2iE_{c}^{2}\sqrt{E_{c}^{2}-m^{2}}F_{V}H_{T}^{*}+2iE_{c}^{2}\sqrt{E_{c}^{2}-m^{2}}H_{\sigma}H_{T}^{*}+2iE_{c}(E_{c}^{2}-m^{2})F_{A}H_{T}^{*}],\nonumber\\

R_{++-+}=&2\sin\theta[\cos\theta+P_{L}-P_{T}^{2}(\cos\theta\cos2\phi+i\sin2\phi)]\nonumber\\
\times&[mE_{c}|F_{V}|^{2}+\frac{E_{c}^{3}}{m}|H_{\sigma}|^{2}+mE_{c}F_{V}H_{\sigma}^{*}+\frac{E_{c}^{3}}{m}H_{\sigma}F_{V}^{*}-m\sqrt{E_{c}^{2}-m^{2}}F_{V}F_{A}^{*}\nonumber\\
-&\frac{E_{c}^{2}\sqrt{E_{c}^{2}-m^{2}}}{m}H_{\sigma}F_{A}^{*}+2iE_{c}^{2}\sqrt{E_{c}^{2}-m^{2}}H_{T}F_{V}^{*}+2iE_{c}^{2}\sqrt{E_{c}^{2}-m^{2}}H_{T}H_{\sigma}^{*}-2iE_{c}(E_{c}^{2}-m^{2})H_{T}F_{A}^{*}],\nonumber\\

R_{+++-}=&2\sin\theta[\cos\theta-P_{L}-P_{T}^{2}(\cos\theta\cos2\phi-i\sin2\phi)]\nonumber\\
\times&[-mE_{c}|F_{V}|^{2}-\frac{E_{c}^{3}}{m}|H_{\sigma}|^{2}-mE_{c}F_{V}H_{\sigma}^{*}-\frac{E_{c}^{3}}{m}H_{\sigma}F_{V}^{*}-m\sqrt{E_{c}^{2}-m^{2}}F_{V}F_{A}^{*}\nonumber\\
-&\frac{E_{c}^{2}\sqrt{E_{c}^{2}-m^{2}}}{m}H_{\sigma}F_{A}^{*}-2iE_{c}^{2}\sqrt{E_{c}^{2}-m^{2}}H_{T}F_{V}^{*}-2iE_{c}^{2}\sqrt{E_{c}^{2}-m^{2}}H_{T}H_{\sigma}^{*}-2iE_{c}(E_{c}^{2}-m^{2})H_{T}F_{A}^{*}],\nonumber\\

R_{--++}=&2\sin^{2}\theta(1-P_{T}^{2}\cos2\phi)[m^{2}|F_{V}|^{2}+\frac{E_{c}^{4}}{m^{2}}|H_{\sigma}|^{2}\nonumber\\
-&4E_{c}^{2}(E_{c}^{2}-m^{2})|H_{T}|^{2}+2E_{c}^{2}{\rm Re}[F_{V}H_{\sigma}^{*}]-4imE_{c}\sqrt{E_{c}^{2}-m^{2}}{\rm Re}[F_{V}H_{T}^{*}]-\frac{4iE_{c}^{3}\sqrt{E_{c}^{2}-m^{2}}}{m}{\rm Re}[H_{\sigma}H_{T}^{*}]],\nonumber\\

R_{-+-+}=&(3+\cos2\theta+4P_{L}\cos\theta+2P_{T}^{2}\sin^{2}\theta\cos2\phi)\nonumber\\
\times&[E_{c}^{2}|F_{V}|^{2}+E_{c}^{2}|H_{\sigma}|^{2}+(E_{c}^{2}-m^{2})|F_{A}|^{2}\nonumber\\
+&2E_{c}^{2}{\rm Re}[F_{V}H_{\sigma}^{*}]-2E_{c}\sqrt{E_{c}^{2}-m^{2}}{\rm Re}[F_{V}F_{A}^{*}]-2E_{c}\sqrt{E_{c}^{2}-m^{2}}{\rm Re}[H_{\sigma}F_{A}^{*}]],\nonumber\\

R_{-++-}=&[2\sin^{2}\theta+P_{T}^{2}(3\cos2\phi+\cos2\theta\cos2\phi-4i\cos\theta\sin2\phi)]\nonumber\\
\times&[E_{c}^{2}|F_{V}|^{2}+E_{c}^{2}|H_{\sigma}|^{2}-(E_{c}^{2}-m^{2})|F_{A}|^{2}\nonumber\\
+&2E_{c}^{2}{\rm Re}[F_{V}H_{\sigma}^{*}]+2iE_{c}\sqrt{E_{c}^{2}-m^{2}}{\rm Im}[F_{V}F_{A}^{*}]+2iE_{c}\sqrt{E_{c}^{2}-m^{2}}{\rm Im}[H_{\sigma}F_{A}^{*}]],\nonumber\\

R_{+--+}=&[2\sin^{2}\theta+P_{T}^{2}(3\cos2\phi+\cos2\theta\cos2\phi+4i\cos\theta\sin2\phi)]\nonumber\\
\times&[E_{c}^{2}|F_{V}|^{2}+E_{c}^{2}|H_{\sigma}|^{2}-(E_{c}^{2}-m^{2})|F_{A}|^{2}\nonumber\\
+&2E_{c}^{2}{\rm Re}[F_{V}H_{\sigma}^{*}]-2i{\rm Im}[F_{V}F_{A}^{*}]E_{c}\sqrt{E_{c}^{2}-m^{2}}-2iE_{c}\sqrt{E_{c}^{2}-m^{2}}{\rm Im}[H_{\sigma}F_{A}^{*}]],\nonumber\\

R_{+-+-}=&[3+\cos2\theta-4P_{L}\cos\theta+2P_{T}^{2}\sin^{2}\theta\cos2\phi]\nonumber\\
\times&[E_{c}^{2}|F_{V}|^{2}+E_{c}^{2}|H_{\sigma}|^{2}+(E_{c}^{2}-m^{2})|F_{A}|^{2}\nonumber\\
+&2E_{c}^{2}{\rm Re}[F_{V}H_{\sigma}^{*}]+2E_{c}\sqrt{E_{c}^{2}-m^{2}}{\rm Re}[F_{V}F_{A}^{*}]+2E_{c}\sqrt{E_{c}^{2}-m^{2}}{\rm Re}[H_{\sigma}F_{A}^{*}]],\nonumber\\

R_{++--}=&2\sin^{2}\theta(1-P_{T}^{2}\cos2\phi)\nonumber\\
\times&[m^{2}|F_{V}|^{2}+\frac{E_{c}^{4}}{m^{2}}|H_{\sigma}|^{2}-4E_{c}^{2}(E_{c}^{2}-m^{2})|H_{T}|^{2}\nonumber\\
+&\frac{4iE_{c}^{3}\sqrt{E_{c}^{2}-m^{2}}}{m}{\rm Re}[H_{\sigma}H_{T}^{*}]+2E_{c}^{2}{\rm Re}[F_{V}H_{\sigma}^{*}]+4imE_{c}\sqrt{E_{c}^{2}-m^{2}}{\rm Re}[F_{V}H_{T}^{*}]],\nonumber\\

R_{---+}=&2\sin\theta[\cos\theta+P_{L}-P_{T}^{2}(\cos\theta\cos2\phi+i\sin2\phi)]\nonumber\\
\times&[mE_{c}|F_{V}|^{2}+\frac{E_{c}^{3}}{m}|H_{\sigma}|^{2}+\frac{E_{c}^{3}}{m}H_{\sigma}F_{V}^{*}+mE_{c}F_{V}H_{\sigma}^{*}-m\sqrt{E_{c}^{2}-m^{2}}F_{V}F_{A}^{*}\nonumber\\
-&\frac{E_{c}^{2}\sqrt{E_{c}^{2}-m^{2}}}{m}H_{\sigma}F_{A}^{*}-2iE_{c}^{2}\sqrt{E_{c}^{2}-m^{2}}H_{T}F_{V}^{*}-2iE_{c}^{2}\sqrt{E_{c}^{2}-m^{2}}H_{T}H_{\sigma}^{*}+2iE_{c}(E_{c}^{2}-m^{2})H_{T}F_{A}^{*}],\nonumber\\

R_{--+-}=&2\sin\theta[\cos\theta-P_{L}-P_{T}^{2}(\cos\theta\cos2\phi-i\sin2\phi)]\nonumber\\
\times&[-mE_{c}|F_{V}|^{2}-\frac{E_{c}^{3}}{m}|H_{\sigma}|^{2}-\frac{E_{c}^{3}}{m}H_{\sigma}F_{V}^{*}-mE_{c}F_{V}H_{\sigma}^{*}-m\sqrt{E_{c}^{2}-m^{2}}F_{V}F_{A}^{*}\nonumber\\
-&\frac{E_{c}^{2}\sqrt{E_{c}^{2}-m^{2}}}{m}H_{\sigma}F_{A}^{*}+2iE_{c}^{2}\sqrt{E_{c}^{2}-m^{2}}H_{T}F_{V}^{*}+2iE_{c}^{2}\sqrt{E_{c}^{2}-m^{2}}H_{T}H_{\sigma}^{*}+2iE_{c}(E_{c}^{2}-m^{2})H_{T}F_{A}^{*}],\nonumber\\

R_{-+--}=&2\sin\theta[\cos\theta+P_{L}-P_{T}^{2}(\cos\theta\cos2\phi-i\sin2\phi)]\nonumber\\
\times&[mE_{c}|F_{V}|^{2}+\frac{E_{c}^{3}}{m}|H_{\sigma}|^{2}+\frac{E_{c}^{3}}{m}H_{\sigma}^{*}F_{V}+mE_{c}F_{V}^{*}H_{\sigma}-m\sqrt{E_{c}^{2}-m^{2}}F_{A}F_{V}^{*}\nonumber\\
-&\frac{E_{c}^{2}\sqrt{E_{c}^{2}-m^{2}}}{m}F_{A}H_{\sigma}^{*}+2iE_{c}^{2}\sqrt{E_{c}^{2}-m^{2}}F_{V}H_{T}^{*}+2iE_{c}^{2}\sqrt{E_{c}^{2}-m^{2}}H_{\sigma}H_{T}^{*}-2iE_{c}(E_{c}^{2}-m^{2})F_{A}H_{T}^{*}],\nonumber\\

R_{+---}=&2\sin\theta[\cos\theta-P_{L}-P_{T}^{2}(\cos\theta\cos2\phi+i\sin2\phi)]\nonumber\\
\times&[-mE_{c}|F_{V}|^{2}-\frac{E_{c}^{3}}{m}|H_{\sigma}|^{2}-\frac{E_{c}^{3}}{m}F_{V}H_{\sigma}^{*}-mE_{c}H_{\sigma}F_{V}^{*}-m\sqrt{E_{c}^{2}-m^{2}}F_{A}F_{V}^{*}\nonumber\\
-&\frac{E_{c}^{2}\sqrt{E_{c}^{2}-m^{2}}}{m}F_{A}H_{\sigma}^{*}-2iE_{c}^{2}\sqrt{E_{c}^{2}-m^{2}}F_{V}H_{T}^{*}-2iE_{c}^{2}\sqrt{E_{c}^{2}-m^{2}}H_{\sigma}H_{T}^{*}-2iE_{c}(E_{c}^{2}-m^{2})F_{A}H_{T}^{*}],\nonumber\\

R_{----}=&2\sin^{2}\theta[1-P_{T}^{2}\cos2\phi]\nonumber\\
\times&[m^{2}|F_{V}|^{2}+\frac{E_{c}^{4}}{m^{2}}|H_{\sigma}|^{2}+4E_{c}^{2}(E_{c}^{2}-m^{2})|H_{T}|^{2}\nonumber\\
+&2E_{c}^{2}{\rm Re}[F_{V}H_{\sigma}^{*}]-4mE_{c}\sqrt{E_{c}^{2}-m^{2}}{\rm Im}[F_{V}H_{T}^{*}]-\frac{4E_{c}^{3}\sqrt{E_{c}^{2}-m^{2}}}{m}{\rm Im}[H_{\sigma}H_{T}^{*}]],
\end{align}
where the symbols $+(-)$ indicates the helicity $\lambda_{1}(\lambda_{2})$ as $+1/2(-1/2)$, $m$ denotes the mass of the hyperon $B$, and $E_c =  M_{J/\psi}/2$ represents half of the center-of-mass energy.
\end{widetext}

\section{DECAY SDM OF HYPERONS}\label{deM}
The decay SDM $T$ for decay process $B\rightarrow B_1 \pi$ can be expressed as
\begin{align}
&T_{++++}=\frac{1}{4}(1+\alpha_{D})(1+\cos\theta),\nonumber\\
&T_{-+++}=\frac{1}{4}(1+\alpha_{D})\sin\theta(\cos\phi-i\sin\phi),\nonumber\\
&T_{+-++}=\frac{1}{4}(i\beta_{D}-\gamma_{D})\sin\theta,\nonumber\\
&T_{++-+}=\frac{1}{4}(1+\alpha_{D})\sin\theta(\cos\phi+i\sin\phi),\nonumber\\
&T_{+++-}=-\frac{1}{4}(i\beta_{D}+\gamma_{D})\sin\theta,\nonumber\\
&T_{--++}=-\frac{1}{4}(i\beta_{D}-\gamma_{D})(1+\cos\theta)(\cos\phi-i\sin\phi),\nonumber\\
&T_{-+-+}=\frac{1}{4}(1+\alpha_{D})(1-\cos\theta),\nonumber\\
&T_{-++-}=-\frac{1}{4}(i\beta_{D}+\gamma_{D})(1-\cos\theta)(\cos\phi-i\sin\phi),\nonumber\\
&T_{+--+}=\frac{1}{4}(i\beta_{D}-\gamma_{D})(1-\cos\theta)(\cos\phi+i\sin\phi),\nonumber\\
&T_{+-+-}=\frac{1}{4}(1-\alpha_{D})(1-\cos\theta),\nonumber\\
&T_{++--}=\frac{1}{4}(i\beta_{D}+\gamma_{D})(1+\cos\theta)(\cos\phi+i\sin\phi),\nonumber\\
&T_{---+}=-\frac{1}{4}(i\beta_{D}-\gamma_{D})\sin\theta,\nonumber\\
&T_{--+-}=-\frac{1}{4}(1-\alpha_{D})\sin\theta(\cos\phi-i\sin\phi),\nonumber\\
&T_{-+--}=\frac{1}{4}(i\beta_{D}+\gamma_{D})\sin\theta,\nonumber\\
&T_{+---}=-\frac{1}{4}(1-\alpha_{D})\sin\theta(\cos\phi+i\sin\phi),\nonumber\\
&T_{----}=\frac{1}{4}(1-\alpha_{D})(1+\cos\theta),
\end{align}
where $\theta$ and $\phi$ denote the polar and azimuthal angles of $B_1$ in $B$ helicity frame, parameters $\alpha_{D}$, $\beta_{D}$, and $\gamma_{D}$, as defined by Eq.~\eqref{abg}, are the decay parameters for $B$. 

\begin{widetext}
\section{ISOSPIN DECOMPOSITION}\label{de}

In this section, we present the correspondence between $CP$ violation observables and the weak and strong phase angles in the decay processes of $\Lambda,~\Sigma$, and $\Xi$ hyperons. This analytical framework was established in Ref.~\cite{donoghue1986hyperon}. In earlier studies, the $\Delta I = 3/2$ partial-wave amplitudes were typically assumed to be small, and only their leading-order contributions were retained. However, with increasing experimental precision, the role of $\Delta I = 3/2$ amplitudes has become increasingly significant~\cite{PhysRevLett.132.101801}. Therefore, in our analysis, we do not expand in powers of $\Delta I = 3/2$ amplitudes. Instead, we perform an expansion in terms of the leading-order of the weak and strong phase angles. This approach is more robust and highlights the potential importance of $\Delta I = 3/2$ contributions. After retaining only the leading-order terms in the $\Delta I = 3/2$ amplitudes, one can recover expressions consistent with those in Refs.~\cite{donoghue1986hyperon,PhysRevD.105.116022}. A numerical estimate of the $\Delta I = 3/2$ partial-wave amplitudes is provided in Appendix~\ref{amplitude_anal}.

The decay amplitudes for $\Lambda\to p\pi^{-}$ are \cite{donoghue1986hyperon}
\begin{align}
A_{S}^{\Lambda\to p\pi}=&-\frac{\sqrt{2}}{\sqrt{3}}S_{11}e^{i\left(\delta_{1}^{S}+\phi_{11}^{S}\right)}+\frac{1}{\sqrt{3}}S_{33}e^{i\left(\delta_{3}^{S}+\phi_{33}^{S}\right)},\label{ppi1}\\
A_{P}^{\Lambda\to p\pi}=&-\frac{\sqrt{2}}{\sqrt{3}}P_{11}e^{i\left(\delta_{1}^{P}+\phi_{11}^{P}\right)}+\frac{1}{\sqrt{3}}P_{33}e^{i\left(\delta_{3}^{P}+\phi_{33}^{P}\right)}.\label{ppi2}
\end{align}
By expanding to leading order in the strong and weak phase angles, we obtain
\begin{align}
\Delta_{CP}^{\Lambda\rightarrow p\pi}=&\frac{2\sqrt{2}\left[P_{11}P_{33}\sin(\delta_{1}^{P}-\delta_{3}^{P})\sin(\phi_{11}^{P}-\phi_{33}^{P})+S_{11}S_{33}\sin(\delta_{1}^{S}-\delta_{3}^{S})\sin(\phi_{11}^{S}-\phi_{33}^{S})\right]}{\left(\sqrt{2}S_{11}-S_{33}\right)^{2}+\left(\sqrt{2}P_{11}-P_{33}\right)^{2}},\\
A_{CP}^{\Lambda\to p\pi}=&\frac{-2S_{11}P_{11}}{\left(\sqrt{2}S_{11}-S_{33}\right)\left(\sqrt{2}P_{11}-P_{33}\right)}\left[\sin\left(\delta_{1}^{P}-\delta_{1}^{S}\right)\sin\left(\phi_{11}^{P}-\phi_{11}^{S}\right)-\frac{1}{\sqrt{2}}\frac{S_{33}}{S_{11}}\sin\left(\delta_{1}^{P}-\delta_{3}^{S}\right)\sin\left(\phi_{11}^{P}-\phi_{33}^{S}\right)\right.\nonumber\\
&\left.-\frac{1}{\sqrt{2}}\frac{P_{33}}{P_{11}}\sin\left(\delta_{3}^{P}-\delta_{1}^{S}\right)\sin\left(\phi_{33}^{P}-\phi_{11}^{S}\right)+\frac{1}{2}\frac{S_{33}P_{33}}{S_{11}P_{11}}\sin\left(\delta_{3}^{P}-\delta_{3}^{S}\right)\sin\left(\phi_{33}^{P}-\phi_{33}^{S}\right)\right]-\Delta_{CP}^{\Lambda\rightarrow p\pi},\\
B_{CP}^{\Lambda\to p\pi}=&\frac{2S_{11}P_{11}}{\left(\sqrt{2}S_{11}-S_{33}\right)\left(\sqrt{2}P_{11}-P_{33}\right)}\left[\sin\left(\phi_{11}^{P}-\phi_{11}^{S}\right)-\frac{1}{\sqrt{2}}\frac{S_{33}}{S_{11}}\sin\left(\phi_{11}^{P}-\phi_{33}^{S}\right)\right.\nonumber\\
&\left.-\frac{1}{\sqrt{2}}\frac{P_{33}}{P_{11}}\sin\left(\phi_{33}^{P}-\phi_{11}^{S}\right)+\frac{1}{2}\frac{S_{33}P_{33}}{S_{11}P_{11}}\sin\left(\phi_{33}^{P}-\phi_{33}^{S}\right)\right],\\
\Delta C^{\Lambda\rightarrow p\pi}=&\frac{2S_{11}P_{11}}{\left(\sqrt{2}S_{11}-S_{33}\right)\left(\sqrt{2}P_{11}-P_{33}\right)}\left[\sin\left(\delta_{1}^{P}-\delta_{1}^{S}\right)-\frac{1}{\sqrt{2}}\frac{S_{33}}{S_{11}}\sin\left(\delta_{1}^{P}-\delta_{3}^{S}\right)\right.\nonumber\\
&\left.-\frac{1}{\sqrt{2}}\frac{P_{33}}{P_{11}}\sin\left(\delta_{3}^{P}-\delta_{1}^{S}\right)+\frac{1}{2}\frac{S_{33}P_{33}}{S_{11}P_{11}}\sin\left(\delta_{3}^{P}-\delta_{3}^{S}\right)\right].
\end{align}

The decay amplitudes for $\Lambda\to n\pi^{0}$ are \cite{donoghue1986hyperon}
\begin{align}
A_{S}^{\Lambda\rightarrow n\pi}=&\frac{1}{\sqrt{3}}S_{11}e^{i\left(\delta_{1}^{S}+\phi_{11}^{S}\right)}+\frac{\sqrt{2}}{\sqrt{3}}S_{33}e^{i\left(\delta_{3}^{S}+\phi_{33}^{S}\right)},\\A_{P}^{\Lambda\rightarrow n\pi}=&\frac{1}{\sqrt{3}}P_{11}e^{i\left(\delta_{1}^{P}+\phi_{11}^{P}\right)}+\frac{\sqrt{2}}{\sqrt{3}}P_{33}e^{i\left(\delta_{3}^{P}+\phi_{33}^{P}\right)},
\end{align}
and we have
\begin{align}
\Delta_{CP}^{\Lambda\rightarrow n\pi}=&-2\sqrt{2}\frac{P_{11}P_{33}\sin\left(\delta_{1}^{P}-\delta_{3}^{P}\right)\sin\left(\phi_{11}^{P}-\phi_{33}^{P}\right)+S_{11}S_{33}\sin\left(\delta_{1}^{S}-\delta_{3}^{S}\right)\sin\left(\phi_{11}^{S}-\phi_{33}^{S}\right)}{\left(S_{11}+\sqrt{2}S_{33}\right)^{2}+\left(P_{11}+\sqrt{2}P_{33}\right)^{2}},\\
A_{CP}^{\Lambda\rightarrow n\pi}=&\frac{-S_{11}P_{11}}{\left(S_{11}+\sqrt{2}S_{33}\right)\left(P_{11}+\sqrt{2}P_{33}\right)}\left[\sin\left(\delta_{1}^{P}-\delta_{1}^{S}\right)\sin\left(\phi_{11}^{P}-\phi_{11}^{S}\right)+\frac{\sqrt{2}S_{33}}{S_{11}}\sin\left(\delta_{1}^{P}-\delta_{3}^{S}\right)\sin\left(\phi_{11}^{P}-\phi_{33}^{S}\right)\right.\nonumber\\
&\left.+\frac{\sqrt{2}P_{33}}{P_{11}}\sin\left(\delta_{3}^{P}-\delta_{1}^{S}\right)\sin\left(\phi_{33}^{P}-\phi_{11}^{S}\right)+2\frac{S_{33}P_{33}}{S_{11}P_{11}}\sin\left(\delta_{3}^{P}-\delta_{3}^{S}\right)\sin\left(\phi_{33}^{P}-\phi_{33}^{S}\right)\right]-\Delta_{\text{CP}}^{\Lambda\rightarrow n\pi},\\
B_{CP}^{\Lambda\rightarrow n\pi}=&\frac{S_{11}P_{11}}{\left(S_{11}+\sqrt{2}S_{33}\right)\left(P_{11}+\sqrt{2}P_{33}\right)}\left[\sin\left(\phi_{11}^{P}-\phi_{11}^{S}\right)+\frac{\sqrt{2}S_{33}}{S_{11}}\sin\left(\phi_{11}^{P}-\phi_{33}^{S}\right)\right.\nonumber\\
&\left.+\frac{\sqrt{2}P_{33}}{P_{11}}\sin\left(\phi_{33}^{P}-\phi_{11}^{S}\right)+2\frac{S_{33}P_{33}}{S_{11}P_{11}}\sin\left(\phi_{33}^{P}-\phi_{33}^{S}\right)\right],\\
\Delta C^{\Lambda\rightarrow n\pi}=&\frac{S_{11}P_{11}}{\left(S_{11}+\sqrt{2}S_{33}\right)\left(P_{11}+\sqrt{2}P_{33}\right)}\left[\sin\left(\delta_{1}^{P}-\delta_{1}^{S}\right)+\sqrt{2}\frac{S_{33}}{S_{11}}\sin\left(\delta_{1}^{P}-\delta_{3}^{S}\right)\right.\nonumber\\
&\left.+\sqrt{2}\frac{P_{33}}{P_{11}}\sin\left(\delta_{3}^{P}-\delta_{1}^{S}\right)+2\frac{S_{33}P_{33}}{S_{11}P_{11}}\sin\left(\delta_{3}^{P}-\delta_{3}^{S}\right)\right].
\end{align}

The decay amplitudes for $\Sigma^{+}\to p\pi^{0}$ are \cite{donoghue1986hyperon}
\begin{align}
A_{S}^{\Sigma^{+}\to p\pi}=&\frac{\sqrt{2}}{3}\left(S_{11}e^{i\phi_{11}^{S}}+\frac{1}{2}S_{31}e^{i\phi_{31}^{S}}\right)e^{i\delta_{1}^{S}}-\frac{\sqrt{2}}{3}\left(S_{13}e^{i\phi_{13}^{S}}-\frac{2\sqrt{2}}{\sqrt{5}}S_{33}e^{i\phi_{33}^{S}}\right)e^{i\delta_{3}^{S}},\\
A_{P}^{\Sigma^{+}\to p\pi}=&\frac{\sqrt{2}}{3}\left(P_{11}e^{i\phi_{11}^{P}}+\frac{1}{2}S_{31}e^{i\phi_{31}^{P}}\right)e^{i\delta_{1}^{P}}-\frac{\sqrt{2}}{3}\left(P_{13}e^{i\phi_{13}^{P}}-\frac{2\sqrt{2}}{\sqrt{5}}P_{33}e^{i\phi_{33}^{P}}\right)e^{i\delta_{3}^{P}}.
\end{align}
Calculating the decay asymmetries, we obtain 
\begin{align}
\Delta_{CP}^{\Sigma^{+}\rightarrow p\pi^{0}}=&2\frac{\bar{S}_{1}\bar{S}_{3}\sin\left(\delta_{1}^{S}-\delta_{3}^{S}\right)\sin\left(\bar{\phi}_{1}^{S}-\bar{\phi}_{3}^{S}\right)+\bar{P}_{1}\bar{P}_{3}\sin\left(\delta_{1}^{P}-\delta_{3}^{P}\right)\sin\left(\bar{\phi}_{1}^{P}-\bar{\phi}_{3}^{P}\right)}{\left(\bar{S}_{1}-\bar{S}_{3}\right)^{2}+\left(\bar{P}_{1}-\bar{P}_{3}\right)^{2}},\\
A_{CP}^{\Sigma^{+}\rightarrow p\pi^{0}}=&\frac{-\bar{P}_{1}\bar{S}_{1}}{\left(\bar{S}_{1}-\bar{S}_{3}\right)\left(\bar{P}_{1}-\bar{P}_{3}\right)}\left[\sin\left(\delta_{1}^{P}-\delta_{1}^{S}\right)\sin\left(\bar{\phi}_{1}^{P}-\bar{\phi}_{1}^{S}\right)-\frac{\bar{S}_{3}}{\bar{S}_{1}}\sin\left(\delta_{1}^{P}-\delta_{3}^{S}\right)\sin\left(\bar{\phi}_{1}^{P}-\bar{\phi}_{3}^{S}\right)\right.\nonumber\\
&\left.-\frac{\bar{P}_{3}}{\bar{P}_{1}}\sin\left(\delta_{3}^{P}-\delta_{1}^{S}\right)\sin\left(\bar{\phi}_{3}^{P}-\bar{\phi}_{1}^{S}\right)+\frac{\bar{S}_{3}\bar{P}_{3}}{\bar{S}_{1}\bar{P}_{1}}\sin\left(\delta_{3}^{P}-\delta_{3}^{S}\right)\sin\left(\bar{\phi}_{3}^{P}-\bar{\phi}_{3}^{S}\right)\right]-\Delta_{CP}^{\Sigma^{+}\rightarrow p\pi^{0}},\\
B_{CP}^{\Sigma^{+}\rightarrow p\pi^{0}}=&\frac{\bar{P}_{1}\bar{S}_{1}}{\left(\bar{S}_{1}-\bar{S}_{3}\right)\left(\bar{P}_{1}-\bar{P}_{3}\right)}\left[\sin\left(\bar{\phi}_{1}^{P}-\bar{\phi}_{1}^{S}\right)-\frac{\bar{S}_{3}}{\bar{S}_{1}}\sin\left(\bar{\phi}_{1}^{P}-\bar{\phi}_{3}^{S}\right)\right.\nonumber\\
&\left.-\frac{\bar{P}_{3}}{\bar{P}_{1}}\sin\left(\bar{\phi}_{3}^{P}-\bar{\phi}_{1}^{S}\right)+\frac{\bar{S}_{3}\bar{P}_{3}}{\bar{S}_{1}\bar{P}_{1}}\sin\left(\bar{\phi}_{3}^{P}-\bar{\phi}_{3}^{S}\right)\right],\\
\Delta C^{\Sigma^{+}\rightarrow p\pi^{0}}=&\frac{\bar{P}_{1}\bar{S}_{1}}{\left(\bar{S}_{1}-\bar{S}_{3}\right)\left(\bar{P}_{1}-\bar{P}_{3}\right)}\left[\sin\left(\delta_{1}^{P}-\delta_{1}^{S}\right)-\frac{\bar{S}_{3}}{\bar{S}_{1}}\sin\left(\delta_{1}^{P}-\delta_{3}^{S}\right)\right.\nonumber\\
&\left.-\frac{\bar{P}_{3}}{\bar{P}_{1}}\sin\left(\delta_{3}^{P}-\delta_{1}^{S}\right)+\frac{\bar{S}_{3}\bar{P}_{3}}{\bar{S}_{1}\bar{P}_{1}}\sin\left(\delta_{3}^{P}-\delta_{3}^{S}\right)\right],
\end{align}
where
\begin{align}
\bar{S}_{1}=&S_{11}+\frac{1}{2}S_{31},~\bar{S}_{3}=S_{13}-2\sqrt{\frac{2}{5}}S_{33},\\
\bar{\phi}_{1}^{S}=&\frac{S_{11}\phi_{11}^{S}+\frac{1}{2}S_{31}\phi_{31}^{S}}{S_{11}+\frac{1}{2}S_{31}},~\bar{\phi}_{3}^{S}=\frac{S_{13}\phi_{13}^{S}-2\sqrt{\frac{2}{5}}S_{33}\phi_{33}^{S}}{S_{13}-2\sqrt{\frac{2}{5}}S_{33}}.
\end{align}
The decay amplitudes for $\Sigma^{+}\rightarrow n\pi^{+}$ are \cite{donoghue1986hyperon}
\begin{align}
A_{S}^{\Sigma^{+}\rightarrow n\pi^{+}}=&\frac{2}{3}\left(S_{11}e^{i\phi_{11}^{S}}+\frac{1}{2}S_{31}e^{i\phi_{31}^{S}}\right)e^{i\delta_{1}^{S}}+\frac{1}{3}\left(S_{13}e^{i\phi_{13}^{S}}-\frac{2\sqrt{2}}{\sqrt{5}}S_{33}e^{i\phi_{33}^{S}}\right)e^{i\delta_{3}^{S}},\\
A_{P}^{\Sigma^{+}\rightarrow n\pi^{+}}=&\frac{2}{3}\left(P_{11}e^{i\phi_{11}^{P}}+\frac{1}{2}S_{31}e^{i\phi_{31}^{P}}\right)e^{i\delta_{1}^{P}}+\frac{1}{3}\left(P_{13}e^{i\phi_{13}^{P}}-\frac{2\sqrt{2}}{\sqrt{5}}P_{33}e^{i\phi_{33}^{P}}\right)e^{i\delta_{3}^{P}},
\end{align}
and we have
\begin{align}
\Delta_{CP}^{\Sigma^{+}\rightarrow n\pi^{+}}=&-4\frac{\bar{S}_{1}\bar{S}_{3}\sin\left(\delta_{1}^{S}-\delta_{3}^{S}\right)\sin\left(\bar{\phi}_{1}^{S}-\bar{\phi}_{3}^{S}\right)+\bar{P}_{1}\bar{P}_{3}\sin\left(\delta_{1}^{P}-\delta_{3}^{P}\right)\sin\left(\bar{\phi}_{1}^{P}-\bar{\phi}_{3}^{P}\right)}{\left(2\bar{S}_{1}+\bar{S}_{3}\right){}^{2}+\left(2\bar{P}_{1}+\bar{P}_{3}\right){}^{2}},\\
A_{CP}^{\Sigma^{+}\rightarrow n\pi^{+}}=&-\frac{4\bar{P}_{1}\bar{S}_{1}}{\left(2\bar{S}_{1}+\bar{S}_{3}\right)\left(2\bar{P}_{1}+\bar{P}_{3}\right)}\left[\sin\left(\delta_{1}^{P}-\delta_{1}^{S}\right)\sin\left(\bar{\phi}_{1}^{P}-\bar{\phi}_{1}^{S}\right)+\frac{\bar{S}_{3}}{2\bar{S}_{1}}\sin\left(\delta_{1}^{P}-\delta_{3}^{S}\right)\sin\left(\bar{\phi}_{1}^{P}-\bar{\phi}_{3}^{S}\right)\right.\nonumber\\
&\left.+\frac{\bar{P}_{3}}{2\bar{P}_{1}}\sin\left(\delta_{3}^{P}-\delta_{1}^{S}\right)\sin\left(\bar{\phi}_{3}^{P}-\bar{\phi}_{1}^{S}\right)+\frac{\bar{S}_{3}\bar{P}_{3}}{4\bar{S}_{1}\bar{P}_{1}}\sin\left(\delta_{3}^{P}-\delta_{3}^{S}\right)\sin\left(\bar{\phi}_{3}^{P}-\bar{\phi}_{3}^{S}\right)\right]-\Delta_{CP}^{\Sigma^{+}\rightarrow n\pi^{+}},\\
B_{CP}^{\Sigma^{+}\rightarrow n\pi^{+}}=&\frac{4\bar{P}_{1}\bar{S}_{1}}{\left(2\bar{S}_{1}+\bar{S}_{3}\right)\left(2\bar{P}_{1}+\bar{P}_{3}\right)}\left[\sin\left(\bar{\phi}_{1}^{P}-\bar{\phi}_{1}^{S}\right)+\frac{\bar{S}_{3}}{2\bar{S}_{1}}\sin\left(\bar{\phi}_{1}^{P}-\bar{\phi}_{3}^{S}\right)\right.\nonumber\\
&\left.+\frac{\bar{P}_{3}}{\bar{P}_{1}}\sin\left(\bar{\phi}_{3}^{P}-\bar{\phi}_{1}^{S}\right)+\frac{\bar{S}_{3}\bar{P}_{3}}{4\bar{S}_{1}\bar{P}_{1}}\sin\left(\bar{\phi}_{3}^{P}-\bar{\phi}_{3}^{S}\right)\right],\\
\Delta C^{\Sigma^{+}\rightarrow n\pi^{+}}=&\frac{4\bar{P}_{1}\bar{S}_{1}}{\left(2\bar{S}_{1}+\bar{S}_{3}\right)\left(2\bar{P}_{1}+\bar{P}_{3}\right)}\left[\sin\left(\delta_{1}^{P}-\delta_{1}^{S}\right)+\frac{\bar{S}_{3}}{2\bar{S}_{1}}\sin\left(\delta_{1}^{P}-\delta_{3}^{S}\right)\right.\nonumber\\
&\left.+\frac{\bar{P}_{3}}{2\bar{P}_{1}}\sin\left(\delta_{3}^{P}-\delta_{1}^{S}\right)+\frac{\bar{S}_{3}\bar{P}_{3}}{4\bar{S}_{1}\bar{P}_{1}}\sin\left(\delta_{3}^{P}-\delta_{3}^{S}\right)\right].
\end{align}
The decay amplitudes for $\Sigma^{-}\rightarrow n\pi^{-}$ are \cite{donoghue1986hyperon}
\begin{align}
A_{S}^{\Sigma^{-}\rightarrow n\pi^{-}}=&\left(S_{13}e^{i\phi_{13}^{S}}+\frac{\sqrt{2}}{\sqrt{5}}S_{33}e^{i\phi_{33}^{S}}\right)e^{i\delta_{3}^{S}},\\
A_{P}^{\Sigma^{-}\rightarrow n\pi^{-}}=&\left(P_{13}e^{i\phi_{13}^{P}}+\frac{\sqrt{2}}{\sqrt{5}}P_{33}e^{i\phi_{33}^{P}}\right)e^{i\delta_{3}^{P}},
\end{align}
and we have
\begin{align}
\Delta_{CP}^{\Sigma^{-}\rightarrow n\pi^{-}}=&0,\\
A_{CP}^{\Sigma^{-}\rightarrow n\pi^{-}}=&\frac{-5\bar{P}_{3}\bar{S}_{3}}{\left(\sqrt{5}\bar{P}_{3}+3\sqrt{2}P_{33}\right)\left(\sqrt{5}\bar{S}_{3}+3\sqrt{2}S_{33}\right)}\nonumber\\
&\times\left[\sin\left(\delta_{3}^{P}-\delta_{3}^{S}\right)\sin\left(\bar{\phi}_{3}^{P}-\bar{\phi}_{3}^{S}\right)+\frac{3\sqrt{2}S_{33}}{\sqrt{5}\bar{S}_{3}}\sin\left(\delta_{3}^{P}-\delta_{3}^{S}\right)\sin\left(\bar{\phi}_{3}^{P}-\phi_{33}^{S}\right)\right.\nonumber\\
&\left.+\frac{3\sqrt{2}P_{33}}{\sqrt{5}\bar{P}_{3}}\sin\left(\delta_{3}^{P}-\delta_{3}^{S}\right)\sin\left(\phi_{33}^{P}-\bar{\phi}_{3}^{S}\right)+\frac{18P_{33}S_{33}}{5\bar{P}_{3}\bar{S}_{3}}\sin\left(\delta_{3}^{P}-\delta_{3}^{S}\right)\sin\left(\phi_{33}^{P}-\phi_{33}^{S}\right)\right],\\
B_{CP}^{\Sigma^{-}\rightarrow n\pi^{-}}=&\frac{5\bar{P}_{3}\bar{S}_{3}}{\left(\sqrt{5}\bar{P}_{3}+3\sqrt{2}P_{33}\right)\left(\sqrt{5}\bar{S}_{3}+3\sqrt{2}S_{33}\right)}\left[\sin\left(\bar{\phi}_{3}^{P}-\bar{\phi}_{3}^{S}\right)\right.\nonumber\\
&\left.+\frac{3\sqrt{2}S_{33}}{\sqrt{5}\bar{S}_{3}}\sin\left(\bar{\phi}_{3}^{P}-\phi_{33}^{S}\right)+\frac{3\sqrt{2}P_{33}}{\sqrt{5}\bar{P}_{3}}\sin\left(\phi_{33}^{P}-\bar{\phi}_{3}^{S}\right)+\frac{18P_{33}S_{33}}{5\bar{P}_{3}\bar{S}_{3}}\sin\left(\phi_{33}^{P}-\phi_{33}^{S}\right)\right],\\
\Delta C^{\Sigma^{-}\rightarrow n\pi^{-}}=&\frac{5\bar{P}_{3}\bar{S}_{3}\sin\left(\delta_{3}^{P}-\delta_{3}^{S}\right)}{\left(\sqrt{5}\bar{P}_{3}+3\sqrt{2}P_{33}\right)\left(\sqrt{5}\bar{S}_{3}+3\sqrt{2}S_{33}\right)}\left[1+\frac{3\sqrt{2}S_{33}}{\sqrt{5}\bar{S}_{3}}+\frac{3\sqrt{2}P_{33}}{\sqrt{5}\bar{P}_{3}}+\frac{18P_{33}S_{33}}{5\bar{P}_{3}\bar{S}_{3}}\right].
\end{align}

The decay amplitudes for $\Xi^{-}\to\Lambda\pi^{-}$ are \cite{donoghue1986hyperon}
\begin{align}
A_{S}^{\Xi^{-}\to\Lambda\pi^{-}}=&\left(S_{12}e^{i\phi_{12}^{S}}+\frac{1}{2}S_{32}e^{i\phi_{32}^{S}}\right)e^{i\delta_{2}^{S}},\\
A_{P}^{\Xi^{-}\to\Lambda\pi^{-}}=&\left(P_{12}e^{i\phi_{12}^{P}}+\frac{1}{2}P_{32}e^{i\phi_{32}^{P}}\right)e^{i\delta_{2}^{P}}.
\end{align}
The discussion goes completely parallel to that for $\Lambda$ decay, and we find
\begin{align}
\Delta_{CP}^{\Xi^{-}\rightarrow\Lambda\pi^{-}}=&0,\\
A_{CP}^{\Xi^{-}\rightarrow\Lambda\pi^{-}}=&\frac{-4S_{12}P_{12}\sin\left(\delta_{2}^{P}-\delta_{2}^{S}\right)}{\left(2S_{12}+S_{32}\right)\left(2P_{12}+P_{32}\right)}\left[\sin\left(\phi_{12}^{P}-\phi_{12}^{S}\right)+\frac{1}{2}\frac{S_{32}}{S_{12}}\sin\left(\phi_{12}^{P}-\phi_{32}^{S}\right)\right.\nonumber\\
&\left.+\frac{1}{2}\frac{P_{32}}{P_{12}}\sin\left(\phi_{32}^{P}-\phi_{12}^{S}\right)+\frac{1}{4}\frac{S_{32}P_{32}}{S_{12}P_{12}}\sin\left(\phi_{32}^{P}-\phi_{32}^{S}\right)\right],\\
B_{CP}^{\Xi^{-}\rightarrow\Lambda\pi^{-}}=&\frac{4S_{12}P_{12}}{\left(2S_{12}+S_{32}\right)\left(2P_{12}+P_{32}\right)}\left[\sin\left(\phi_{12}^{P}-\phi_{12}^{S}\right)+\frac{1}{2}\frac{S_{32}}{S_{12}}\sin\left(\phi_{12}^{P}-\phi_{32}^{S}\right)\right.\nonumber\\
&\left.+\frac{1}{2}\frac{P_{32}}{P_{12}}\sin\left(\phi_{32}^{P}-\phi_{12}^{S}\right)+\frac{1}{4}\frac{S_{32}P_{32}}{S_{12}P_{12}}\sin\left(\phi_{32}^{P}-\phi_{32}^{S}\right)\right],\label{XimB}\\
\Delta C^{\Xi^{-}\rightarrow\Lambda\pi^{-}}=&\tan\left(\delta_{2}^{P}-\delta_{2}^{S}\right),\label{XimC}
\end{align}
while for $\Xi^{0}\to\Lambda\pi^{0}$
\begin{align}
A_{S}^{\Xi^{0}\to\Lambda\pi^{0}}=&\frac{1}{\sqrt{2}}\left(S_{12}e^{i\phi_{12}^{S}}-S_{32}e^{i\phi_{32}^{S}}\right)e^{i\delta_{2}^{S}},\\
A_{P}^{\Xi^{0}\to\Lambda\pi^{0}}=&\frac{1}{\sqrt{2}}\left(P_{12}e^{i\phi_{12}^{P}}-P_{32}e^{i\phi_{32}^{P}}\right)e^{i\delta_{2}^{P}}.
\end{align}
And we have
\begin{align}
\Delta_{CP}^{\Xi^{0}\rightarrow\Lambda\pi^{0}}=&0,\\
A_{CP}^{\Xi^{0}\rightarrow\Lambda\pi^{0}}=&\frac{-S_{12}P_{12}\sin\left(\delta_{2}^{P}-\delta_{2}^{S}\right)}{\left(S_{12}-S_{32}\right)\left(P_{12}-P_{32}\right)}\left[\sin\left(\phi_{12}^{P}-\phi_{12}^{S}\right)-\frac{S_{32}}{S_{12}}\sin\left(\phi_{12}^{P}-\phi_{32}^{S}\right)\right.\nonumber\\
&\left.-\frac{P_{32}}{P_{12}}\sin\left(\phi_{32}^{P}-\phi_{12}^{S}\right)+\frac{S_{32}P_{32}}{S_{12}P_{12}}\sin\left(\phi_{32}^{P}-\phi_{32}^{S}\right)\right],\\
B_{CP}^{\Xi^{0}\rightarrow\Lambda\pi^{0}}=&\frac{S_{12}P_{12}}{\left(S_{12}-S_{32}\right)\left(P_{12}-P_{32}\right)}\left[\sin\left(\phi_{12}^{P}-\phi_{12}^{S}\right)-\frac{S_{32}}{S_{12}}\sin\left(\phi_{12}^{P}-\phi_{32}^{S}\right)\right.\nonumber\\
&\left.-\frac{P_{32}}{P_{12}}\sin\left(\phi_{32}^{P}-\phi_{12}^{S}\right)+\frac{S_{32}P_{32}}{S_{12}P_{12}}\sin\left(\phi_{32}^{P}-\phi_{32}^{S}\right)\right],\label{Xi0B}\\
\Delta C ^{\Xi^{0}\rightarrow\Lambda\pi^{0}}= & \tan\left(\delta_{2}^{P}-\delta_{2}^{S}\right).\label{Xi0C}
\end{align}

\section{PARTIAL-WAVE AMPLITUDE ANALYSIS}\label{amplitude_anal}

In this section, we present a numerical analysis of the contributions from different partial-wave amplitudes for $\Lambda$ and $\Xi$ decays. While a similar analysis has been performed in Ref.~\cite{PhysRevD.105.116022}, we offer an alternative approach to this investigation.

For the decay process $B_1\to B_2\pi$, the decay width is given by
\begin{align}
\Gamma=\frac{|\vec{p}|}{8\pi m_1^2}\frac{1}{2}\sum_{\lambda_1\lambda_2}|\mathcal{M}_{\lambda_1,\lambda_2}|^2,
\end{align}
where $m_1$ is the mass of the $B_1$ hyperon, $|\vec{p}|$ is the momentum of the $B_2$ baryon in the $B_1$ rest frame, and $\lambda_1$ and $\lambda_2$ are the helicity of $B_1$ and $B_2$, respectively. The decay amplitude $\mathcal{M}$ can be expressed as~\cite{PDG} 
\begin{align}
\mathcal{M}_{\lambda_1\lambda_2}=G_Fm_\pi^2\bar{u}_{B_1}(A-B\gamma_5)u_{B_2},
\end{align}
where $G_F=1.166\times10^{-5}\text{ GeV}^{-2}$ is the Fermi coupling parameter, and $m_\pi=(m_{\pi^+}+m_{\pi^-}+m_{\pi^0})/3$ denotes the averaged pion mass. The parameters $A$ and $B$ are related to the S- and P-wave amplitudes via
\begin{align}
&A_{S}=A,\nonumber\\
&A_{P}=\frac{|\vec{p}|}{(E_2+m_2)}B,
\end{align}
with $m_2$ being the $B_2$ baryon mass and $E_2$ is its energy in the $B_1$ rest frame. Thus, we obtain
\begin{align}\label{Gamma}
\Gamma&=\frac{|\vec{p}|(E_2+m_2)}{4\pi m_1}G_F^2 m_\pi^4(A^2+\frac{|\vec p|^2}{(E_2+m_2)^2}B^2)\nonumber\\
&=\frac{|\vec{p}|(E_2+m_2)}{4\pi m_1}G_F^2 m_\pi^4(A_S^2+A_P^2).
\end{align}
The partial wave amplitudes for the decay processes $\Lambda \to p\pi^-$ and $\Lambda \to n\pi^0$ are provided in Appendix~\ref{de}. Under isospin conservation, both $\Lambda \to p\pi^-$ and $\Lambda \to n\pi^0$ share identical final-state strong interaction phases $\delta_i^L$. However, the mass difference between the decay products introduces momentum‐dependent corrections to these phase~\cite{PhysRevD.105.116022}. To account for this, we denote the corrected strong phases in the $\Lambda \to n\pi^0$ decay as $\tilde{\delta}_i^L$. We summarize the phase differences between these processes in Table~\ref{phaseDiff}.

\begin{table*}[b]
\centering
\caption{\label{phaseDiff}Values of the $N-\pi$ scattering phase shifts $\delta_{2I}^L$ relevant for $\Lambda$ decays from~\cite{HOFERICHTER20161}.}
\renewcommand{\arraystretch}{1.5}
\begin{ruledtabular}
\begin{tabular}{llcccc}
\multicolumn{1}{c}{}&\multicolumn{1}{c}{$\delta_1^S~(^\circ)$}  &\multicolumn{1}{c}{$\delta_3^S~(^\circ)$}&\multicolumn{1}{c}{$\delta_1^P~(^\circ)$}&\multicolumn{1}{c}{$\delta_3^P~(^\circ)$}\\\hline

$\Lambda\to p\pi^-$&   $6.39\pm0.09$    &  $-4.46\pm0.07$    & $-0.77\pm0.07$   &  $-0.71\pm0.03$   \\
$\Lambda\to n\pi^0$&  $6.58\pm0.10$ & $-4.66\pm0.07$&$-0.80\pm0.08$& $-0.77\pm0.04$\\                                                                          
\end{tabular}
\end{ruledtabular}
\end{table*}

Since the weak phase angle $\phi_{ij}$, which represents $CP$ violation, is very small, it is safe to neglect it in the analysis of this section. Then, the decay widths and asymmetry parameters for $\Lambda\to p \pi^-$ and $\Lambda\to n\pi^0$ are given by
\begin{align}
\Gamma_{\Lambda\rightarrow p\pi}=&\frac{\left|\vec{p}\right|\left(E_{p}+m_{p}\right)}{4\pi m_{\Lambda}}\frac{G_{F}^2m_{\pi}^{4}}{3}\left[2S_{11}^{2}+2P_{11}^{2}+P_{33}^{2}+S_{33}^{2}-2\sqrt{2}P_{11}P_{33}\cos(\delta_{1}^{P}-\delta_{3}^{P})-2\sqrt{2}S_{11}S_{33}\cos(\delta_{1}^{S}-\delta_{3}^{S})\right],\\
\Gamma_{\Lambda\rightarrow n\pi}=&\frac{\left|\vec{\tilde{p}}\right|\left(\tilde{E}_{p}+m_{n}\right)}{4\pi m_{\Lambda}}\frac{G_{F}^2m_{\pi}^{4}}{3}\left[S_{11}^{2}+P_{11}^{2}+2P_{33}^{2}+2S_{33}^{2}+2\sqrt{2}P_{11}P_{33}\cos(\tilde{\delta}_{1}^{P}-\tilde{\delta}_{3}^{P})+2\sqrt{2}S_{11}S_{33}\cos(\tilde{\delta}_{1}^{S}-\tilde{\delta}_{3}^{S})\right],\\\alpha_{\Lambda\rightarrow p\pi}=&\frac{2\left[2P_{11}S_{11}\cos(\delta_{1}^{P}-\delta_{1}^{S})-\sqrt{2}P_{11}S_{33}\cos(\delta_{1}^{P}-\delta_{3}^{S})-\sqrt{2}P_{33}S_{11}\cos(\delta_{1}^{S}-\delta_{3}^{P})+P_{33}S_{33}\cos(\delta_{3}^{P}-\delta_{3}^{S})\right]}{2S_{11}^{2}+2P_{11}^{2}+P_{33}^{2}+S_{33}^{2}-2\sqrt{2}P_{11}P_{33}\cos(\delta_{1}^{P}-\delta_{3}^{P})-2\sqrt{2}S_{11}S_{33}\cos(\delta_{1}^{S}-\delta_{3}^{S})},\\
\alpha_{\Lambda\rightarrow n\pi}=&\frac{2\left(P_{11}S_{11}\cos(\tilde{\delta}_{1}^{P}-\tilde{\delta}_{1}^{S})+\sqrt{2}P_{11}S_{33}\cos(\tilde{\delta}_{1}^{P}-\tilde{\delta}_{3}^{S})+\sqrt{2}P_{33}S_{11}\cos(\tilde{\delta}_{1}^{S}-\tilde{\delta}_{3}^{P})+2P_{33}S_{33}\cos(\tilde{\delta}_{3}^{P}-\tilde{\delta}_{3}^{S})\right)}{S_{11}^{2}+P_{11}^{2}+2P_{33}^{2}+2S_{33}^{2}+2\sqrt{2}P_{11}P_{33}\cos(\tilde{\delta}_{1}^{P}-\tilde{\delta}_{3}^{P})+2\sqrt{2}S_{11}S_{33}\cos(\tilde{\delta}_{1}^{S}-\tilde{\delta}_{3}^{S})},
\end{align}
where
\begin{align}
\left|\vec{p}\right|=&\frac{1}{2m_{\Lambda}}\sqrt{(m_{\Lambda}-m_{p}+m_{\pi^{-}})(m_{\Lambda}-m_{p}-m_{\pi^{-}})(m_{\Lambda}+m_{p}-m_{\pi^{-}})(m_{\Lambda}+m_{p}+m_{\pi^{-}})},\\\left|\vec{\tilde{p}}\right|=&\frac{1}{2m_{\Lambda}}\sqrt{(m_{\Lambda}-m_{n}+m_{\pi^{0}})(m_{\Lambda}-m_{n}-m_{\pi^{0}})(m_{\Lambda}+m_{n}-m_{\pi^{0}})(m_{\Lambda}+m_{n}+m_{\pi^{0}})},\\E_{p}=&\frac{m_{\Lambda}^{2}+m_{p}^{2}-m_{\pi^{-}}^{2}}{2m_{\Lambda}},\\\tilde{E}_{p}=&\frac{m_{\Lambda}^{2}+m_{n}^{2}-m_{\pi^{0}}^{2}}{2m_{\Lambda}}.
\end{align}
Using experimental inputs~\cite{PDG,PhysRevLett.132.101801} and theoretical constraints, we numerically determine the partial wave amplitudes $A_S$ and $A_P$. The coupled nonlinear nature of these equations prevents analytical solutions, but reliable numerical results can be obtained using standard computational tools such as \textit{Mathematica} or \textsc{ROOT}.

The quadratic form of the equations yields multiple numerical solutions. To identify the physically meaningful solution branch, we impose the following constraints on the partial wave amplitudes:
\begin{align}
&S_{11}>0,\quad|S_{11}|>|S_{33}|,\quad|P_{11}|>|P_{33}|,\quad\left|A_{S}\right|_{\Lambda\rightarrow p \pi}>\left|A_{p}\right|_{\Lambda\rightarrow p \pi}.
\end{align}
These conditions ensure that the phase of $S_{11}$ is zero, the decay parameter $\gamma_{\Lambda\rightarrow p \pi}$ is positive~\cite{PDG}, and the $\Delta I=1/2$ channel dominates over the $\Delta I=3/2$ channel—a relationship widely confirmed in weak decays.

Table~\ref{phaseDiff} reveals that the final-state strong interaction phase differences are small, yielding the approximation:
\begin{align}
\cos(\delta_i^L-\delta_j^{L^{\prime}})\approx\cos(\tilde\delta_i^L-\tilde\delta_j^{L^{\prime}})\approx 1.
\end{align}
The impact of strong phases on isospin analysis is limited. If we neglect them, the error in partial wave amplitudes can be determined analytically. Taking the target partial wave set as  $\bm{p}=\{S_{11},S_{33},P_{11},P_{33}\}$ and the parameter set as  $\bm{x}=\{\Gamma,\tilde\Gamma,\alpha,\tilde\alpha\}$, we obtain
\begin{align}
J_p=\frac{\partial {\bm p}}{\partial\bm x}=(\frac{\partial\bm x}{\partial \bm p})^{-1}=(J_x)^{-1},
\end{align}
and the error for partial wave amplitude reduces to 
\begin{align}
\sigma_{p,i}=\sqrt{J_{p,ij}\sigma_{x,jj}(J_p)^{T}_{ji}},
\end{align}
where $\sigma_{x,jj}=\sigma^2_{x,j}$ is the error matrix of the parameter $\bm x$. The mean and error of parameters are valued as~\cite{PDG,PhysRevLett.132.101801}
\begin{align}
&\Gamma_{\Lambda\rightarrow p\pi}=\left(1.612\pm0.013\right)\times10^{-15}\text{ GeV},\\
&\Gamma_{\Lambda\rightarrow n\pi}=\left(0.903\pm0.013\right)\times10^{-15}\text{ GeV},\\
&\alpha_{\Lambda\rightarrow p\pi}=0.764\pm0.008,\\
&\alpha_{\Lambda\rightarrow n\pi}=0.670\pm0.009.
\end{align}
Then we can determine both the magnitudes and uncertainties of the partial wave amplitudes when strong phases neglected.

When retaining the final-state strong interaction phases, we perform a straightforward Monte Carlo analysis to estimate uncertainties. Each input parameter, whether derived from experimental measurements or theoretical calculations, is treated as an independent Gaussian random variable characterized by its mean and standard deviation. We generate $10^5$ pseudodata samples for each parameter from these Gaussian distributions. Using these pseudodata sets, we numerically solve the nonlinear equations and obtain all possible solutions. We then perform an error analysis separately for each set of solutions.

The numerical results for the partial-wave amplitudes from both methods are summarized in Table~\ref{SandP}.

Similarly, for the decay $\Xi^-\to\Lambda\pi^-$ and $\Xi^0\to\Lambda\pi^0$, the decay widths and asymmetry parameters are given by
\begin{align}
&\Gamma_{\Xi^{-}\rightarrow\Lambda\pi^{-}}=\frac{\left|\vec{p}\right|\left(E_{\Lambda}+m_{\Lambda}\right)}{4\pi m_{\Xi^{-}}}\frac{G_{F}^2m_{\pi}^{4}}{4}\left((2P_{12}+P_{32})^{2}+(2S_{12}+S_{32})^{2}\right),\\
&\Gamma_{\Xi^{0}\rightarrow\Lambda\pi^{0}}=\frac{\left|\vec{\tilde{p}}\right|\left(\tilde{E}_{\Lambda}+m_{\Lambda}\right)}{4\pi m_{\Xi^{0}}}\frac{G_{F}^2m_{\pi}^{4}}{2}\left((P_{12}-P_{32}){}^{2}+(S_{12}-S_{32}){}^{2}\right),\\
&\alpha_{\Xi^{-}\rightarrow\Lambda\pi^{-}}=\frac{2(2P_{12}+P_{32})(2S_{12}+S_{32})\cos(\delta_{2}^{P}-\delta_{2}^{S})}{(2P_{12}+P_{32})^{2}+(2S_{12}+S_{32})^{2}},\\
&\alpha_{\Xi^{0}\rightarrow\Lambda\pi^{0}}=\frac{2(P_{12}-P_{32})(S_{12}-S_{32})\cos(\tilde{\delta}_{2}^{P}-\tilde{\delta}_{2}^{S})}{(P_{12}-P_{32}){}^{2}+(S_{12}-S_{32}){}^{2}},
\end{align}
with the kinematic quantities
\begin{align}
\left|\vec{p}\right|=&\frac{\sqrt{(m_{\Xi^{-}}-m_{\Lambda}-m_{\pi^{-}})(m_{\Xi^{-}}-m_{\Lambda}+m_{\pi^{-}})(m_{\Xi^{-}}+m_{\Lambda}-m_{\pi^{-}})(m_{\Xi^{-}}+m_{\Lambda}+m_{\pi^{-}})}}{2m_{\Xi^{-}}},\\\left|\vec{\tilde{p}}\right|=&\frac{\sqrt{(m_{\Xi^{0}}-m_{\Lambda}-m_{\pi^{0}})(m_{\Xi^{0}}-m_{\Lambda}+m_{\pi^{0}})(m_{\Xi^{0}}+m_{\Lambda}-m_{\pi^{0}})(m_{\Xi^{0}}+m_{\Lambda}+m_{\pi^{0}})}}{2m_{\Xi^{0}}},\\E_{\Lambda}=&\frac{m_{\Xi^{-}}^{2}+m_{\Lambda}^{2}-m_{\pi^{-}}^{2}}{2m_{\Xi^{-}}},\\\tilde{E}_{\Lambda}=&\frac{m_{\Xi^{0}}^{2}+m_{\Lambda}^{2}-m_{\pi^{0}}^{2}}{2m_{\Xi^{0}}}.
\end{align}

Following experimental constraints~\cite{PDG,BESIII:2021ypr,PhysRevLett.132.101801}, we fix the parameter values and their uncertainties as
\begin{align}
&\Gamma_{\Xi^{-}\rightarrow\Lambda\pi^{-}}=\left(4.01133\pm0.00014\right)\times10^{-15}\text{ GeV},\\
&\Gamma_{\Xi^{0}\rightarrow\Lambda\pi^{0}}=\left(2.25885\pm0.00027\right)\times10^{-15}\text{ GeV},\\
&\alpha_{\Xi^{-}\rightarrow\Lambda\pi^{-}}=-0.367\pm0.004,\\
&\alpha_{\Xi^{0}\rightarrow\Lambda\pi^{0}}=-0.3750\pm0.0034,
\end{align}
while the final-state strong phases are determined to be~\cite{BESIII:2021ypr,PhysRevLett.132.101801}
\begin{align}
&\delta_{2}^{P}-\delta_{2}^{S}=\left(1.89\pm1.15\right)^{\circ},~\text{for}~\Xi^-\to\Lambda\pi^-~\text{decay},\\
&\tilde{\delta}_{2}^{P}-\tilde{\delta}_{2}^{S}=\left(-0.74\pm0.97\right)^{\circ},~\text{for}~\Xi^0\to\Lambda\pi^0~\text{decay}.
\end{align}
Following the same computational framework, we estimate the magnitudes and uncertainties of various partial wave amplitudes in $\Xi$ decays, as presented in Table~\ref{SandP}.
\end{widetext}
\bibliography{apssamp.bib}
\end{document}